\documentclass[10pt]{article}
\usepackage{amsmath,amssymb,amsthm}
\usepackage{epsfig}
\usepackage{graphicx}
\usepackage{rotating}

\title{Ice Spiral Patterns on the Ocean Surface}
\author{Z. Zong$^{1,2,3}$,  A. Ludu$^{4}$
	\\
	\\ \small{1. School of Shipbuilding Engineering, Dalian University of Technology}
	\\ \small{2. Collaborative Centre of Advanced Ships and Deepwater Engineering}
	\\ \small{3. Liaoning Deepwater Floating Structure Engineering Technology Lab}
	\\ \small{4. Department of Mathematics, Embry-Riddle Aeronautical University} \\ \small{Daytona Beach, FL, USA}
	\\}

\begin{document}
	
	\maketitle 
	{\let\thefootnote\relax\footnotetext{{\em Emails}:
			zongzhi@dlut.edu.cn, ludua@erau.edu}}

\begin{abstract}
\noindent

We investigate a new two-dimensional compressible Navier-Stokes hydrodynamic model design to explain and study large scale ice swirls formation at the surface of the ocean. The linearized model generates a basis of Bessel solutions from where various types of spiral patterns can be generated and their evolution and stability in time analyzed. By restricting the nonlinear system of equations to its  quadratic terms  we obtain swirl solutions emphasizing logarithmic spiral geometry. The resulting solutions are analyzed and validated using three mathematical approaches: one predicting the formation of patterns as Townes solitary modes, another approach mapping the nonlinear system into a sine-Gordon equation, and a third approach uses a series expansion.  Pure radial, azimuthal and spiral modes are obtained from the fully nonlinear equations. Combinations of multiple-spiral solutions are also obtained, matching the experimental observations. The  nonlinear stability of the spiral patterns is analyzed by Arnold's convexity method, and the Hamiltonian of the solutions is plotted versus some order parameters showing the existence of geometric phase transitions.

\end{abstract}
%\noindent{{\bf Keywords:} }

\section{Introduction}
\label{sec.intro}

The rapid decline of summer ice extent that has occurred in the Arctic Ocean over recent years has prompted a surge of research activity and, in particular, the role of sea-ice morphology, has been increasingly recognized. This is especially relevant in the context of climate change as the resulting ice melting and proliferation of open water promote further wave growth over increasing fetches, thus allowing
long waves to propagate larger distances into the ice field. Of particular interest is the marginal ice zone which is the fragmented part of the ice cover closest to the open ocean and, as such, it is a very dynamic region strongly affected by ocean currents. The formation of sea-ice fragments increases
the further penetration and damage of the ice cover, so it is highly important to understand the dynamics of these fragments on the ocean surface. In this respect, a lot of research was dedicated to linear and nonlinear ocean waves propagating through fragmented sea-ice \cite{oceanice,ice1,ice2,ice3,ice7,ice9}. 

Sea-ice can directly collide with marine structures, \cite{ice7}, and induce wave scattering or attenuation, \cite{coll1,coll2}, thus indirectly varying the hydrodynamic response of passing vessels  and offshore platform.  To safely manage the navigation and functioning of such structures, and to develop accurate models for their interaction with sea-ice, it is important to be able to predict the distribution and morphology of sea-ice fragments.

\begin{figure}
	\centering
		\includegraphics[width=5cm,height=5cm]{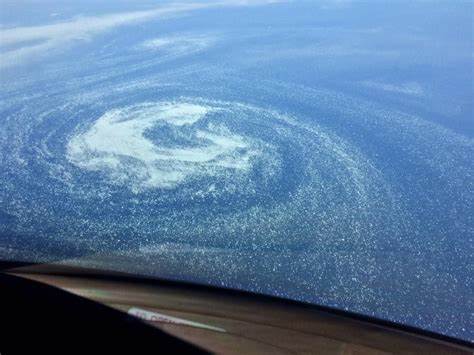} 
	\caption{Large sea-ice swirl observed in the arctic ocean from airplane \cite{1,2}.}
	\label{fignews}
\end{figure}

The mechanism for the formation of swirl patterns observed mostly in arctic ocean, see Figs. \ref{fignews},\ref{figswirl}, \cite{1,2}, has not yet been fully understood or modeled. There is little doubt, from the observational data available, that these huge scale phenomenon is  associated with the ocean currents and possibly wind and waves \cite{moreswirls}. Nevertheless, given the very slow time scale, it is unlikely for these spiral to be generated or enhanced by Coriolis force. A complete model should consider local viscosity generated by collisions of ice fragments, and subsequently the local phase transitions, clustering effect, the vertical motion of water and the elevation waves effect, the wind, and a certain probability distribution for the ice fragments size and shapes.   Such a complete theory, taking into consideration all these components and forces into account, and put their relative importance into perspective is needed, but such a theory is not yet available. 
There is a considerable body of research on the interaction between waves, \cite{ice3,oceanice,ice7,ice9},  solitons and sea-ice, \cite{ice1,ice2}, and quasi-granular aggregate models for the sea-ice \cite{ocean4}. The most natural explanation regarding such sea-ice swirl structure is the formation of a wave pattern, which either remains stationary, or at least quasi-stationary, in a frame of reference
rotating around its center at a proper angular speed. Similar waves of patterns structures were used to explain the spiral galaxies formation \cite{galaxy1}. In this manuscript we favor the point of view that the sea-ice can maintain a density wave through water currents
interaction. This density wave provides a spiral density field which underlies the
observable concentration of sea-ice. In this way, an observable swirl
pattern can be maintained over the whole structure. In this paper we demonstrate the formation and stability of such ice swirls. We introduce a two-dimensional model describing large space-time scale spiral patterns of sea-ice fragments floating on water, like the ones observed in arctic ocean, \cite{1,2}, see for example Fig. \ref{fignews}.

The paper is organized as follows: in section 2 we  introduce a 2-dimensional two-phases compressible fluid model governed on mass and momentum conservation of water plus ice. 
In section 3 we expand the density and velocity fields in a two-scale series, controlled by two smallness parameters and the system of equations is linearized and solved exactly under initial and boundary conditions and the sea-ice swirls solutions are obtained  and discussed. In section 4  we analyze the time evolution of various types of linearized solutions and spirals and their stability. In section 5 we introduce limiting situations for the ice patterns, namely the pure radial motion, the azimuthal (rotational) motion, and the spiral patterns. In section 6, following a qualitative discussion on the structure of the nonlinear system and its solutions, we find solutions  by using three approaches: mapping the nonlinear system into a sine-Gordon equation, using an iterative procedure of partial differential operators, and using a quadratic truncation. In section 7 we study the nonlinear stability of the spiral solutions.

\section{Inviscid two-fluid model}
\label{sec.2fluidmodel}

In our two-dimensional hydrodynamic model the water (density $\rho_w$) is assumed incompressible  and inviscid, and  elevation waves at the water surface are neglected.  Multiple studies of both linear, \cite{coll3}, and nonlinear, \cite{ice1,ice2,ice3},
surface waves propagating through fragmented sea-ice show that the attenuation by scattering,  damping, multiple wave reflections and ice viscosity is most effective  for floe
configuration representing a good compromise between ice concentration and ice fragmentation, combination which represents the typical distribution in the ice swirl patterns.
Given the relative short attenuation length of such ocean waves through sea-ice fragments, \cite{ice2,ice3,ice7}, and the long life time of the observed ice swirls, it seems natural to neglect the influence of ocean waves in modeling the dynamics of ice swirl formation, while ocean currents and winds represent to main contributions.

We consider that the ice fragments (density $\rho_i$) have average size much smaller than the characteristic size of the pattern, so we can assume an almost uniform size, shape, and volume distribution among fragments, and almost constant ice draft (denoted  $T_0$), except special local situation at the center of the swirl. We model the mixture between water and floating sea-ice fragments as a compressible two-dimensional fluid in the surfactant approximation. The flow takes places in the $(x, y)$-plane with  $z$-axis oriented upwards, and $z=0$ being the water surface. The two-fluid system is considered in isothermal and isobaric equilibrium,  without substantial amounts of phase transitions between ice and water. Other higher order of approximation effects, like vertical displacement of ice fragments induced by ocean waves, by wind or by adjacent ice fragments collision, ice elasticity, or strong collisions of ice fragments followed by local ice melting will be considered in a subsequent model.

For the extended ice swirl formations the average value of the water flow is relatively small  $|\vec{v}| \sim 1 \div 10$ mm/s,  \cite{icetrav}, while the sea-ice fragments horizontal size can range between $10 \div 50$ m  \cite{ocean4}. In this conditions the  Reynolds number ranges Re$\sim 10^4 \div 10^7$, the Froude number ranges
Fr$\sim 10^{-6} \div 10^{-4}$ and Euler number is of the order of the unit. Consequently,  the system is governed by the law of mass conservation for the mixture ice-water and by the compressible Cauchy linear momentum equation \cite{5}
\begin{equation}\label{eq6}
\frac{\partial \rho}{\partial t}+\nabla \cdot (\rho \vec{v})=0,
\end{equation}
\begin{equation}\label{eq7}
\frac{\partial (\rho \vec{v})}{\partial t}+\nabla \cdot (\rho \vec{v} \otimes \vec{v})=-\nabla P+\nabla \cdot \mathcal{T}+\vec{f},
\end{equation}
where $\vec{v}$ is the flow field, $\rho$ is the density, $P$ is the total pressure, the symbol $\otimes$ is the direct tensor product, $\vec{f}$ is the volume density of external forces, and $\mathcal{T}$ is the second order symmetric viscous stress tensor
\begin{equation}\label{eq8}
\mathcal{T}^{ki}=\eta \biggl( \frac{\partial v^i}{\partial x_k}+\frac{\partial v^k}{\partial x_i} \biggr),
\end{equation}
and $\eta$ is the dynamical viscosity. The high Reynolds number value and the low Froude number value for the ice swirl patterns suggest a sub-critical regime in which Newtonian viscosity can be neglected in the first approximation. Indeed, from the Navier-Stokes Eq. (\ref{eq7}) written in dimensionless form 
%\frac{D \vec{v}}{Dt}=-\hbox{Eu}\nabla P+\frac{1}{\hbox{Re}}\triangle %\vec{v}+\frac{1}{\hbox{Fr}^2}\vec{f},
is known that the Poiseuille number Poi$=$Eu$\cdot$Re$\sim PL/(v \eta)$ controls the importance of the viscosity term, where $L$ is the typical length scale of the system. The pressure of water upon the
ice fragments can be evaluated by considering water uniformly pushing with
a Stokes drag force upon the submerged regions of the ice fragment. For a regular
prismatic ice fragment, for example,  it results Poi $\sim 0.05$ which means that it is sufficient to consider viscosity in the second order of approximation, for the linearized equations. If for some reason the ice fragments tend to cluster (for example at  large impact velocity) the friction between the mutual fragments can trigger local melting followed by re-solidification, which generate a local increase in the effective viscosity of the mixture. In such a case the floating ice mixture can be considered at large scale as a non-Newtonian fluid, specifically a share thickening (dilatant) mixture where viscosity increases with the rate of shear strain. In other words, the ice fragments and water can transition to coagulation, forming larger ice blocks. In this case the water-ice mixture can be modeled as a Bingham type of fluid with a viscous shear-thinning power flow \cite{3}. Such situation occurs towards the central bulge of the spiral. In the following, we elaborate on a basic model for inviscid flow. 

For the two-dimensional mixture of water sea-ice fragments
we introduce the Ice Fraction Function (IFF) as the surface density of ice vs.
surface, namely the ratio of ice in a unit surface
\begin{equation}\label{eq1}
\Phi(x,y,t)=\frac{A_i}{A}, \ \ 0\le \Phi \le 1,
\end{equation}
where $A_i$ is the area occupied by brush ice in the unit area, and $A$ is the unit
area under consideration. Correspondingly, the area occupied by free surface
water is $A_w =A (1-\Phi)$.

Similarly to the introduction of fractional volume scalar field in air-water-ice-structure CFD modeling, \cite{ice7}, we introduce the density of the water-ice mixture by
\begin{equation}\label{eq3}
\rho(x,y,t)=\Phi \rho_i +(1-\Phi) \rho_w .
\end{equation}
The movement of sea-ice and ocean currents (wind stress is neglected here) are combined in the pressure term in our model. Since we neglect the deformation of sea surface, the  oceanic currents act on ice by ocean–ice interfacial stress which is related to the velocity difference between the ice movement and surface ocean currents. The momentum transfer is usually from the ice to the ocean, \cite{ocean3}, case in which the pressure is given by the ice pressure on water through its buoyancy. Such pressure distribution acting on the sides of the ice fragments is considered positive because acts towards water. However, the pressure can become slightly negative if the momentum transfer is from water towards the ice \cite{ocean1}. Usually in sea-ice modeling, \cite{ocean2}, the pressure is taken as the weighted sum  of pressures of the two phases
\begin{equation}\label{eq4}
P(x,y,t)=\Phi P_i +(1-\Phi)P_w , 
%\simeq \Phi P_i ,
\end{equation}
where $P_i, P_w$ are the ice and water pressure, respectively.  Neglecting
the dynamic effects because the motion is very slow, the pressure on an ice fragment
is approximately equal to its static pressure. We can express the ice pressure
per unit of length of ice horizontal perimeter by
\begin{equation}\label{eq5}
P_i =\rho_w gT.
\end{equation}
As we mentioned above, we assume this draft to be in average
independent of the ice fragment, so we treat it as a constant $T = T_0$.
In the following we use the notation $\vec{v}=u \vec{e}_{r}+v \vec{e}_{\theta}$ for the velocity field, and $\hat{i}, \hat{j}, \vec{e}_r$ and $\vec{e}_{\theta}$ for the unit vectors,  correspondingly. The derivatives with respect to $t,x,y,r,\theta$ are labeled with the corresponding letter subscript, while all other subscripts in this text (like for example $i,w$ for ice and water) do not represent derivatives.

In the following, by introducing Eq. (\ref{eq3}) in Eq. (\ref{eq6}) with $\eta=0$ and expressed in polar coordinates $(r, \theta)$, we obtain a nonlinear differential equation for the mass conservation, in terms of IFF and velocity components
\begin{equation}\label{eq9}
(1 -a \Phi)  \biggl( u_r +\frac{u}{r} + \frac{v_{\theta}}{r} \biggr) -a \biggl( \Phi_t +u \Phi_r   + \frac{v \Phi_{\theta}}{r} \biggr)=0,
\end{equation}
where $a= 1-\rho_i / \rho_w$ is a constant. 
By implementing Eqs. (\ref{eq3}-\ref{eq5}) in Eq. (\ref{eq7}) without the viscosity term, we obtain an equation for each polar component of the law of momentum conservation
\begin{equation}\label{eq10}
(1 -a \Phi) \biggl( u_t  +u u_r +\frac{v u_{\theta}}{r} -\frac{v^2}{r} \biggr) -a \biggl( u \Phi_t  + u^2 \Phi_r + \frac{uv\Phi_{\theta}}{r} \biggr) =\biggl( \frac{P_{w}}{\rho_{w}}-gT_0 \biggr) \Phi_r ,
\end{equation}
for radial component, and 
\begin{equation}\label{eq11}
(1 -a \Phi) \biggl( v_t +\frac{uv}{r}+uv_r +\frac{v v_{\theta}}{r} \biggr)  -a \biggl( v \Phi_t + u v \Phi_r +\frac{v^2 \Phi_{\theta}}{r}\biggr) =\biggl( \frac{P_{w}}{\rho_{w}}-gT_0 \biggr)  \frac{\Phi_{\theta}}{r},
\end{equation}
for the azimuthal one. This system can be written in a more compact form using the notation $\Psi=(1-a \Phi)$ and $b=P_{w}/\rho_{w}-gT_0$
\begin{equation}\label{eq9b}
\Psi_t + (\Psi u)_r   + \frac{(\Psi v)_{\theta}}{r} +\frac{\Psi u}{r}=0,
\end{equation}
\begin{equation}\label{eq10b}
(\Psi u)_t +u(\Psi u)_r +\frac{v (\Psi u)_{\theta}}{r}-\frac{\Psi v^2}{r}+\frac{b}{a}\Psi_r =0,
\end{equation}
\begin{equation}\label{eq11b}
(\Psi v)_t +u(\Psi v)_r +\frac{v (\Psi v)_{\theta}}{r}+\frac{\Psi uv}{r}+\frac{b}{ar}\Psi_{\theta}=0.
\end{equation}
The system Eqs. (\ref{eq9}-\ref{eq11}) , or Eqs. (\ref{eq9b}-\ref{eq11b}), describes the dynamics of the three fields $\Phi$ (or $\Psi$), $u$ and $v$ depending on time $t\ge 0$ and polar coordinates $r\in (0,\infty ), \theta \in (-\infty, \infty )$.  In the following, we  solve the system Eqs. (\ref{eq9}-\ref{eq11b}) under various boundary and initial conditions and discuss exact and asymptotic solutions.

%%%%%%%%%%%%%%%%%%%%%%%%%%%%%%%%%%%%%%%%%%%%%%%%%%%%%%%%%%%%%%%%%%%%%%%%%%%%%%
%%%%%%%%%%%%%%%%%%%%%%%%%%%%%%%%%%%%%%%%%%%%%%%%%%%%%%%%%%%%%%%%%%%%%%%%%%%%%%
%%%%%%%%%%%%%%%%%%%%%%%%%%%%%%%%%%%%%%%%%%%%%%%%%%%%%%%%%%%%%%%%%%%%%%%%%%%%%%
%%%%%%%%%%%%%%%%%%%%%%%%%%%%%%%%%%%%%%%%%%%%%%%%%%%%%%%%%%%%%%%%%%%%%%%%%%%%%%

\section{Linearized model}
\label{sec.lin.sol}

In this section we obtain exact solutions for the linear approximation of the conservative two-fluid model. A quick look at the linearization of Eqs. (\ref{eq10b}, \ref{eq11b})
$$
(\Psi u)_t +\frac{b}{a}\Psi_r =0, \ \  (\Psi v)_t +\frac{b}{ar}\Psi_{\theta}=0,
$$
can give already a hint about the geometry of the solution's patterns. For any of the coefficients of the double Fourier series expansion in time ($i \omega t$), and in the polar angle ($i m \theta$)  of the solution, we have
$$
\frac{u}{v}=-\frac{i}{m}\frac{r \Psi_{r}}{\Psi}.
$$
The left hand side describes the equation  for the stream lines expressed in polar coordinates by  $r=R(\theta)$  
$$
\frac{u}{v}=\frac{R_{\theta}}{R}.
$$
If the flow follows, for example, a logarithmic spiral ($R=R_0 \exp{g_{0} \theta}$) we obtain
$$
\Psi=e^{i(m g_0 \ln r+m \theta+\omega t)},
$$
which indeed represents a one-arm logarithmic spiral pattern in the ice density.

It is natural to ask whether distribution of sea-ice in the form of a rotating spiral is in a state of stable equilibrium. Instability can take the form of a warping of the spiral shape or dispersing the spiral pattern into a more uniform distribution of sea-ice. To obtain the states of equilibrium we consider the linearization of Eqs. (\ref{eq9}-\ref{eq11}) and we expect these linear solutions to generate wave type of equations and dispersion laws from where we can calculate the phase and group velocity, the latest being responsible for the dynamics of sea-ice patterns. Then, as in all stability problems, we consider the small perturbations from the equilibrium state. In the linear approximation spiral patterns are possible only through the effect of interacting waves of considerably different scales \cite{6}. Consequently, we choose  to use  the method of multi-scale expansion in which we express the dependent variables in asymptotic formal series with respect to two small, dimensionless parameters $\varepsilon$ and $\delta$ as follows
$$
\Phi=\Phi^0+\sum_{n=1}^{\infty}\varepsilon^{n} \Phi^{n} (r,\theta,t),
$$
\begin{equation}\label{eq12}
u=\sum_{n=1}^{\infty}\delta^{n} u^{n} (r,\theta,t), \ \ v=
%v^{0}+
\sum_{n=1}^{\infty}\delta^{n} v^{n} (r,\theta,t),
\end{equation}
where we are making the hypothesis that the zero orders of all quantities are time and position independent, and the velocity field has negligible zero order terms. These assumptions are validated by geophysical observations,  \cite{1,2,icetrav,ocean4,ice2,ice3,ice7}, showing that such huge spiral sea-ice structures rotate slow and rather like a rigid pattern.
% shape where the radial flow  always appears to be one order of magnitude smaller than the tangential velocity of the mixture. 

In order to separate different physical space-time scales in the mixed flow  we are using the method of scaled parameters, \cite{odebook}, for  the independent variables to $r = R\zeta$ and $t = \Theta \tau$, using $\zeta, \tau$ as dimensionless independent variables. Since the water flow is faster than the motion of ice fragments we expect the higher order  corrections in the water flow field to dominate the corrections of the IFF density field, which implies $\varepsilon < \delta$, so we can choose $\mathcal{O}(\epsilon)=\mathcal{O}(\delta^2)$ without any loss of generality.   The ice fragments move together with the spiral pattern with the group velocity $v_{gr}$. It is simple to demonstrate that for this system the group velocity is always smaller than the phase  velocity $\hbox{max}||\vec{v}||\simeq v_{gr} < v_{ph}$.  Indeed, a simple estimation for the phase velocity for the pattern waves can be given by $v_{ph} \simeq \sqrt{\partial P / \partial \rho } $. From Eqs. (\ref{eq3}-\ref{eq5}) it results $v_{ph} \simeq \sqrt{g T_0} $. For regular size ice fragments floating freely we can assume $T_0 \sim 10 - 100$ m which sets the phase velocity  in the range  $v_{ph} \sim 20$ m/s $\gg v_{gr} \simeq 0.01$ m/s. On the other hand, the phase velocity obtained from the linearized system of equations is of order $v_{ph}=\omega/k  \sim R/\Theta$.  From these observations it results $ |u| \sim |v| \sim v_{gr}< v_{ph}$. From this scale hierarchy it results the balancing of same order types of terms
\begin{equation}\label{eq13}
v_{gr} \sim \mathcal{O}(u) \sim \mathcal{O}(v) \sim \frac{\varepsilon R}{\delta \Theta}\sim   \frac{\varepsilon \Theta v_{ph}^2}{\delta R}. 
\end{equation}

By using the material conditions of our model Eqs. (\ref{eq3}-\ref{eq5}) in Eqs. (\ref{eq9}-\ref{eq11}) in absence of viscosity, we can re-write the system  Eqs. (\ref{eq9}-\ref{eq11}) in the form
\begin{equation}\label{eq14}
-a \Phi_t - a u \Phi_r +(1-a \Phi) u_r+(1 -a \Phi) \frac{u}{r}+(1 -a \Phi) \frac{v_{\theta}}{r}-a \frac{v\Phi_{\theta}}{r}=0,
\end{equation}
$$
(1-a \Phi) u_t-a u \Phi_t +(1-a \Phi) u u_r-a u^2 \Phi_r +(1-a \Phi)\frac{v u_{\theta}}{r}
$$
\begin{equation}\label{eq15}
-a \frac{uv\Phi_{\theta}}{r}-(1-a \Phi) \frac{v^2}{r}=\biggl( \frac{P_{w}}{\rho_{w}}-g T_0 \biggr) \Phi_r ,
\end{equation}
$$
(1-a \Phi) v_t -a v \Phi_t +(1-a \Phi) u v_r -(1-a \Phi)  \frac{uv}{r} -a \rho u v \Phi_r
$$
\begin{equation}\label{eq16}
+(1-a\Phi )\frac{v v_{\theta}}{r}-a \rho \frac{v^2 \Phi_{\theta}}{r}=\biggl(\frac{P_{w}}{\rho_{w}}- g T_0 \biggr) \frac{\Phi_{\theta}}{r}.
\end{equation} 
In the first order in $\varepsilon, \delta$ Eqs. (\ref{eq14}-\ref{eq16})  become
\begin{equation}\label{eq17}
\frac{\varepsilon R}{\delta \Theta}\Phi_{\tau}^{1}-\biggl(\frac{1}{a}-\Phi_0 \biggr) \biggl(  \frac{u^1 }{\zeta} +  \frac{ v_{\theta}^{1}}{\zeta}+  u^{1}_{\zeta}\biggr)=0,
\end{equation}
$$
(1-a \Phi^0 ) u^{1}_{\tau}=\frac{\varepsilon \Theta}{\delta R}\biggl(\frac{P_{w}}{\rho_{w}}- g T_0 \biggr)  \Phi^{1}_{\zeta},
$$
\begin{equation}\label{eq18}
(1-a \Phi^0 ) v^{1}_{\tau}=\frac{\varepsilon \Theta}{\delta R}\biggl(\frac{P_{w}}{\rho_{w}}- g T_0 \biggr) \frac{\Phi^{1}_{\theta}}{\zeta},
\end{equation}
where we kept the terms in the same order of magnitude according to Eq. (\ref{eq13}). 

In order to  linearize the system Eqs. (\ref{eq17}-\ref{eq18}) we note that $\theta$ and $t$ are cyclic
variables since they do not appear explicitly in the equations, so there are two constants of motion related to these two variables. Moreover, the equations are invariant to Galilean rotations around $\theta$ so the solution should depend on these variables only through phase expressions $ m \theta+ \omega t$, $m$ being integer in order to satisfy periodicity. From the rotational symmetry of the equations we expect that some of the solutions exhibit the same symmetry and provide rotational patterns in rotation. Any steady linearized solution in the first order has uniform density distribution with any divergence free velocity field.  From the integrability conditions in the first order, $\Phi^{1} _{r \theta}=\Phi^{1}_{\theta r}$, and by using  Eqs. (\ref{eq17}-\ref{eq18}) we re-obtain the vorticity $\vec{w}=\nabla \times \vec{v}=\vec{k} w$ conservation law in the first order
$$
\frac{\partial }{\partial t} [ u^{1}_{\theta}-(r v^{1})_{r} ]=-r w^{1}_t =0,
$$
conservation validates our choice for the smallness orders in the model.  We differentiate Eqs (\ref{eq17}) with respect to $\tau$, and differentiate Eqs. (\ref{eq18}) with respect to $\zeta$ and $\theta$ respectively, and plug the results into the $\tau$ derivative of Eqs. (\ref{eq17}). Eq. (\ref{eq17}) decouples the density function $\Phi^1$ from velocity field and, back in dimensional variables, takes the form of a wave equation in cylindrical coordinates
\begin{equation}\label{eq20}
\Phi^{1}_{tt}-c^2 \biggl( \Phi^{1}_{rr}+\frac{\Phi^{1}_{r}}{r}+\frac{\Phi^{1}_{\theta \theta}}{r^2} \biggr)=0,
\end{equation}
where the constant in front of the second term
\begin{equation}\label{eq19}
c^2=\frac{P_{w}-\rho_{w} g T_0}{\rho_w - \rho_i}=\frac{\frac{P_w}{\rho_w}-g T_0}{a},
\end{equation}
if positive, it can be interpreted as the phase velocity of the system's linear waves. Eq. (\ref{eq20}) can be solved by separation of variables, and the general solution has the form
\begin{equation}\label{eq21}
\Phi^1 =  Z_m \biggl( \frac{\omega r}{c} \biggr) e^{i (m \theta +\omega t)}, 
\end{equation}
where $\omega=\omega_r + i \omega_i$ is a separation parameter with its real part $\omega_r$ representing the angular frequency of the waves, $m$ is an arbitrary integer, and $Z_m$ is a linear combination of Bessel functions of first or second kind of integer order $m$.
In principle the solution must be regular at the origin $r=0$, which would limit the general solutions of Eq. (\ref{eq21}) to be only Bessel functions of the first kind $J_m ( \omega r/c)$. However, the real sea-ice swirl structure cannot be covered by the linearized model at its very central part where the ice always clusters into a more or less solid fragment, and hence the mixture cannot be treated as an idea compressible fluid. We therefore admit a singularity at $r = 0$ in the solution Eq. (\ref{eq21}) and request the solutions to describe a small radius $r_0$ ice core at the center of the swirl, see arrow 2 in Fig. \ref{fignews}. 
If $\omega$ is real and the water pressure exceeds the ice pressure ($P_w > \rho_w g T_0 $) the argument of the Bessel functions is real, the oscillations of the radial function, and hence spiral shapes are allowed in the asymptotic region. If the water pressure is less than 
the ice pressure, we have $c^2 <0$, the radial part of the solution becomes modified Bessel function, and oscillations are absent.
The solution for IFF approaches  asymptotically a uniform distributed value $\Phi_{\infty}\in [0,1]$ towards infinity, definitely outside the perimeter $L$ of the swirl, see see arrow 1 in Fig. \ref{fignews}. In other words the parameter $\omega$ will be determined from suitable chosen boundary conditions for $\lim\limits_{r\rightarrow 0}\Phi \sim \Phi(r_0)= 1$ and $\lim\limits_{r\rightarrow \infty}\Phi \sim \Phi(L)=\Phi_{\infty}$, where $L$ will be a suitable chosen parameter describing the radius of the swirl. These boundary conditions for our solutions request a discrete spectrum for $\omega$. Physically, this means that we are setting an exact solution as a representation of the swirl  part of the mixture, and we leave the dynamics near the center, where ice concentration is larger and maybe even three-dimensional, to adjust itself to almost any requirement of the
swirl part. We also expect that all the perturbations would decay to zero at infinity in a smooth 
manner.

With these comments being said, the the radial part of the solutions of Eq. (\ref{eq20}) will be denoted generically by any Bessel function $Z_m$ and specified in more detail when is needed. These solutions involve damped wave oscillating evolution in the radial direction of the IFF, since in the far asymptotic range $r\rightarrow \infty$  the Bessel functions approach $1/\sqrt{r}$ times a periodic oscillating function in argument. This asymptotic behavior results in the coupling of the radial periodicity of the asymptotic expression of the Bessel solutions with the natural periodicity in the azimuthal angle and with time, hence generating solutions  $\Phi^1$ with circular  pattern symmetries, including spirals
\begin{equation}\label{eq32345}
\Phi^{1}|_{r\to\infty}\sim \sqrt{ \frac{c}{2 \pi \omega_r r}} \hbox{Exp}\biggl[ i \biggl( \frac{\omega_r r}{c}-\frac{m \pi}{2}+m \theta+\omega_r t \biggr) \biggr].
\end{equation} 
This asymptotic solution represents an Archimedean with $m$ arms, and these are trailing spiral arms if $\omega_r /c <0$, and  leading spiral arms if  $\omega_r /c >0$ \cite{galaxy1}. The spirals represent stable modes if $\omega_i >0$. The Bessel solutions are bounded on $[r_0,\infty)$ matching the necessary range for $\Phi \in [0,1]$, and they always decreases asymptotically to zero towards $r$ approaching infinity, meaning that far away from the center of the spiral the sea-ice degenerates in constant and uniform distribution, rather controlled at this point by random dynamics. 

In order to build the general solution of the linearized system Eqs. (\ref{eq17}-\ref{eq18}) we use linear combinations of Eq. (\ref{eq21}) summed over all ranges of parameters $m,\omega$   
\begin{equation}\label{eq23}
\Phi^{1} (r, \theta, t)=\sum_{m=-\infty}^{\infty}\int_{-\infty}^{\infty} C_{m, \omega} Z_{m} \biggl( \frac{\omega}{c} r \biggr) e^{i(m \theta+ \omega_r t)} d\omega.
\end{equation}
These integral form generates solutions for given initial conditions, boundary conditions, or regularity conditions at $r=r_0$ and $r=L$ or even $r \rightarrow \infty$. Actually, even this general solution 
can describe spirals when the amplitude of the Bessel functions varies slowly with the radial distance, and if the phase varies quickly (large $\omega_r$ values). Indeed using the integral representation for Bessel functions, \cite{5}, we have
\begin{equation}\label{eq43}
\Phi^{1} (r, \theta, t)=\sum_{m=-\infty}^{\infty}\int_{-\infty}^{\infty} \int_{-\pi}^{\pi} C_{m, \omega, \tau} \hbox{Exp} \biggl[ i \biggl( \frac{\sin \tau \omega_r}{c} r  +m \theta+ \omega_r t) \biggr) \biggr]
d\tau d\omega,
\end{equation}
which is a linear combination of Archimedean spirals of equations
$$
r=-\frac{c}{\omega_r \sin \tau} (m \theta+\omega_r t)
$$

We built solutions within a disk of radius $L \gg r_0$. Physically, further away from the boundary of this disk $L$ we have only a uniform and sparse mixture of water with less and less ice, so we can impose the boundary condition $\Phi^{1}(L, \theta, t)=0$. Under this assumption, and inspired by Eq. (\ref{eq23}),  we can consider that the solution in Eqs. (\ref{eq21}) form, for each $m$, an orthogonal system of functions labeled by $n$, complete over the space of continuous functions defined inside the disk $L$, and for $t>0$  in the form
\begin{equation}\label{eq24}
\Phi^{1}_{m,n}= H_{m} \biggl( \frac{\xi_{m,n}}{L} r \biggr) e^{i(m \theta+ \frac{c \xi_{m,n}}{L} t)},
\end{equation}
where $\xi_{m,n}$ is the root of order $n\ge 1$ of the Hankel function of the first kind and order $m\ge 0$, i.e. $H_{m}(\xi_{m,n})=0$.  The order of the Hankel function is not relevant for the geometry of the spiral. We chose Hankel functions over other types of Bessel functions because these ones have the desired asymptotic behavior to generate spirals. The completeness of this orthogonal system of Bessel functions reduces the integral Eq. (\ref{eq23}) over frequencies to a sum over the Bessel function roots, Eq. (\ref{eq24}).  

With this basis of functions we can solve the initial condition $\Phi^{1}(r, \theta, 0)=\Phi^{1}_{0}(r, \theta)$ problem for the system Eqs. (\ref{eq17}-\ref{eq18}), by expanding the initial condition function in the  a double Fourier and Fourier-Bessel series  and thus obtain the continuous solution  in the form
$$
\Phi^1 (r,\theta,t)=\frac{2}{L^2}\sum_{m=-\infty}^{\infty}\sum_{n=1}^{\infty} \frac{\int_{0}^{L}\int_{0}^{2 \pi} s H_{m}\biggl( \frac{\xi_{m,n}}{L}s\biggr) \Phi^{1}_{0} (s, \phi) e^{-im\phi }ds d\phi}{H_{m+1}^{2}(\xi_{m,n})} \times
$$
\begin{equation}\label{eq25}
\times H_{m}\biggl( \frac{\xi_{m,n}}{L}r \biggr) \hbox{Exp}\biggl[i \biggl( m \theta +\frac{c \xi_{m,n} }{L}t \biggr) \biggr].
\end{equation}
The basis of functions Eq. (\ref{eq24}) built based on the boundary condition $\Phi^{1}(L, \theta, t)=0$ plus the initial condition  provide the general solution for the IFF density field $\Phi^{1}$ in the first order linearized case.

From Eqs. (\ref{eq18}) and (\ref{eq25}) we  obtain the linearized velocity field 
\begin{equation}\label{eq26}
u^1 (r, \theta, t)= \frac{\epsilon \biggl( \frac{P_w}{\rho_w} -g T_0 \biggr)}{\delta(1-a\Phi^0)}\int\limits_{0}^{t} \Phi^{1}_{r} (r,\theta,s) ds, 
\end{equation}
\begin{equation}\label{eq27}
v^1 (r, \theta, t)= \frac{\epsilon \biggl( \frac{P_w}{\rho_w} -g T_0 \biggr)}{\delta r (1-a\Phi^0)}\int\limits_{0}^{t} \Phi^{1}_{\theta} (r,\theta,s) ds.
\end{equation}
All the partial waves in Eq. (\ref{eq24}) have linear dispersion relations with $\omega_{radial} =c k$ and $\omega_{azimuthal} =c \xi_{m,n} k /( m L)$, so the solution Eq. (\ref{eq25}) represents a linear combination of different rotational waves  in the $\theta$ direction, and same type of waves in the radial direction. For large values of $r$,  because of the asymptotic form of the Hankel functions, the azimuthal frequencies tend to be the same for any type of partial wave, $\xi_{m,n}\rightarrow m \pi /L$ so $\omega_{azimuthal} \rightarrow c \pi /(2L)$. That means that for large  $r$ the solutions of  Eqs. (\ref{eq17}-\ref{eq18}) the solution tends to become coherent in space-time, describing large-scale collective and coherent patterns of ice and water, including various number of arms spirals.

A solution can also describe a partial wave, with given $m,n$, and then the IFF field has the form
\begin{equation}\label{eq28}
\Phi^{1}(r,\theta,t)=C H_{m}\biggl( \frac{\xi_{m,n}}{L}r \biggr) \hbox{Exp} \biggl[ i\biggl(m \theta +\frac{c \xi_{m,n}}{L}t \biggr) \biggr].
\end{equation}
The velocity field associated with this density field is given by
\begin{equation}\label{eq29}
u^{1}(r, \theta, t)=-\frac{C \epsilon \biggl( \frac{P_w}{\rho_w} -g T_0 \biggr)}{2 c \delta(1-a\Phi^0)} \biggl[ H_{m-1}\biggl( \frac{\xi_{m,n}}{L}r \biggr) +H_{m+1}\biggl( \frac{\xi_{m,n}}{L}r \biggr) \biggr] e^{i (m \theta +\frac{c \xi_{m,n}}{L}t)},
\end{equation}
and
\begin{equation}\label{eq30}
v^{1}(r, \theta, t)=- \frac{iC\epsilon \biggl( \frac{P_w}{\rho_w} -g T_0 \biggr)}{c \delta (1-a\Phi^0)}  \frac{H_{m}\biggl( \frac{\xi_{m,n}}{L}r \biggr)}{\frac{m \xi_{m,n}r}{L}} e^{i (m \theta +\frac{c \xi_{m,n}}{L}t)}.
\end{equation}
The properties of the general solution Eqs. (\ref{eq25}-\ref{eq27}) are in agreement with all the properties of observed patterns of ice. It is straightforward to check the IFF function fulfills the linear integrability condition, namely $-r w_z = u_{\theta} -(r v)_{r}=$constant. We also note that the only way to have steady motion solution is to cancel the coefficient in front of the time exponential. The only Bessel function fulfilling the case is for $m=0, n=1$   which, when implemented in the first order solution it reduces it to zero.

%%%%%%%%%%%%%%%%%%%%%%%%%%%%%%%%%%%%%%%%%%%%%%%%%%%%%%%%%%%%%%%%%%%%%%%%%%%%%%
%%%%%%%%%%%%%%%%%%%%%%%%%%%%%%%%%%%%%%%%%%%%%%%%%%%%%%%%%%%%%%%%%%%%%%%%%%%%%%
%%%%%%%%%%%%%%%%%%%%%%%%%%%%%%%%%%%%%%%%%%%%%%%%%%%%%%%%%%%%%%%%%%%%%%%%%%%%%%
%%%%%%%%%%%%%%%%%%%%%%%%%%%%%%%%%%%%%%%%%%%%%%%%%%%%%%%%%%%%%%%%%%%%%%%%%%%%%%

\section{Time evolution of the Archimedean spiral}
\label{secTimeevol}

In this section we study the time evolution of some particular initial conditions for ice distribution in order to establish what types of patterns generate long time stable solutions and are more likely to develop and maintain in the sea. The most interesting pattern is the Archimedean spiral described in polar coordinates by the equation $a r + m \theta +\psi =0$, valid for any real $a$, any integer $m$, and any real parameter $\psi$ between $0$ and $2 \pi$. The Archimedean spiral has constant pitch $2 \pi m/a$, that is the radial distance
between successive turns. The integer $m$ describes the number of arms of the spiral, and the last term describes a rotation of  angle $\psi$. If, for example, $\psi=c a t$, the spiral rotates in time like a rigid pattern with constant angular speed $ ac$. Moreover, any function defined in the plane which depends on space coordinates and time only through the variable $\chi=a r + m \theta +v_0 t$
has contour levels with Archimedean spiral shape. Of course the function chosen in this study must be continuous and periodic in $\chi$, otherwise we will have multi-valued representation of the plane. Example of such function can be sine/cosine trigonometric functions, Bessel functions, hyperbolic  functions, etc. We consider that a field defined on the plane has contour levels following Archimedean spiral if it can be written as $\Phi(r, \theta, t)=g(r) P(a r + m \theta +v_0 t)$ with $g$ a decreasing function approaching $0$ at infinity, and $P$ a continuous periodic function.

In the following, we discuss what are the initial conditions that can generate solutions with Archimedean pattern shape, and also the linear stability in time of such solutions. The radial extension of such observed spiral is very large (tens to hundred of Km) so in our calculation we can take at different stages the limit $L \gg R$ and even $L \rightarrow \infty$ which are  realistic choices on one hand, and useful shortcuts in calculations since the integral over radius can be approximated with a Fourier transforms. The time scale of the phenomenon is also very large since the spiral pattern was observed moving very slow and persisting long time. Since in the absence of storms or strong sea currents, the distribution of ice fragments can be considered random, which means the IFF function is a constant. The sudden formation of such an Archimedean spiral pattern represents an interesting phenomenon of large scale collective behavior of ice fragments and water. Such coherent stable structures 
are possible only through a nonlinear dynamics which allows many scales interaction. Nevertheless, the linear solutions obtained in the previous section form a basis of complete functions, so we can expand any hypothetical  nonlinear solution in series of the linear solutions and manage the series coefficients  to assure stability of some nature.

Let us consider a standard Archimedean spiral $\Phi_{0}=\cos(a_0 r +m_0 \theta)$ as initial condition. The solution in Eq. (\ref{eq25}) will retain only  the term with $m=m_0$. In order to obtain the coefficients of the remaining series over $n$ we have to integrate terms of the general form
\begin{equation}\label{eq31}
T_{m_0 ,n}=\hbox{Re }\biggl[\int_{0}^{L} s e^{i a_0 s} H_{m_0}\biggl( \frac{\xi_{m_0 ,n}}{L}s \biggr) ds \bigg].
\end{equation}
Such integrals are exactly the coefficients of the Fourier-Bessel series in $H_{m_0}$ for the initial condition
$$
\cos (a_0 r)= \hbox{Re} \biggl[\sum_{n=1}^{\infty} T_{m_0 , n} H_{m_0 } \biggl(\frac{\xi_{m_0 , n} r}{L} \biggr) \biggr] .
$$
The solution for this initial condition, Eq. (\ref{eq25}), becomes
\begin{equation}\label{eq32}
\Phi^{1} =\frac{4 \pi}{L^2} \hbox{Re} \biggl[ e^{i m_0 \theta} \sum_{n=1}^{\infty} T_{m_0 ,n}H_{m_0}\biggl( \frac{\xi_{m_0 ,n}}{L} r \biggr) e^{\frac{ic\xi_{m_0 ,n}}{L}t} \biggr] .
\end{equation}
In the limit $L$ a very large number, and taking into account only the real part
of the exponential involved, we can approximate the domain of integration with
the positive real semi-axis, and for the even cosine function the numerator in
Eq. (\ref{eq31}) can be approximated with
$$
\hbox{Re} \biggl[\int_{0}^{L} s e^{i a_0 s} H_{m_0} \biggl( \frac{\xi_{m_0 ,n}}{L} s \biggr) ds \biggr]=\frac{L}{2 \xi_{m_0 ,n}}\hbox{Re} \biggl[ \frac{\partial}{\partial \zeta} \hat{H}_{m_0}(\zeta) \biggr]_{\zeta=\frac{a_0 L}{\xi_{m_0 ,n}}} ,
$$
where $\hat{H}_{m_0}(\zeta)$ is the Fourier transform of the corresponding Bessel
function. This Fourier transform is a polynomial of order $m_0$ in $\zeta$, with support only within the unit symmetric interval $\zeta \in [-1,1]$ and zero in the rest. Consequently, its derivative becomes
the sum of two delta-Dirac distribution, out of which only $\delta(\zeta -a_0 L/\xi_{m_0 ,n} )$ matters since $\zeta>0$. It means that in the sum over $n$ in Eq. (\ref{eq32}) only the terms having $n$ given by the solutions of the transcendental equation $a_0 L=\xi_{m_0 ,n}$ have the dominant contribution, and let us denote this solution for $n$ by $n_0$. The rest of the terms with $n\neq n_0$ can be neglected, especially for very large values of $t$ since their contribution is densely chopped by the high frequency oscillations. The resulting solution for Eq. (\ref{eq32}) has the  form
\begin{equation}\label{eq33}
\Phi^{1}= \hbox{Re} \biggl[ C H_{m_0}\biggl( \frac{\xi_{m_0 ,n_0}}{L} r \biggr) e^{i (m_0 \theta +c a_0 t)} \biggr]+\mathcal{O}\biggl(\frac{1}{a_0 ct} \biggr),
\end{equation}
which in the asymptotic limit of very large distance, where the Bessel function can be approximated with cosine  we have
$$
\Phi^{1} \rightarrow  e^{i(a_0 r+ m_0 \theta +c a_0 t)} +\mathcal{O}\biggl( \max \biggl\{ \frac{1}{a_0 ct}, \frac{L}{r} \biggr\} \biggr),
$$
which means that any partial wave initial condition will be linearly stable in time, and it will keep its shape. Nevertheless, such solutions are nonphysical because the spiral extends to infinity and does not decay and smoothly connect to the surrounding water. We need to multiply such initial solutions with a decaying factor with respect to $r$.

\subsection{Time evolution of partial waves}
\label{subsecTimePartial}

The linear stability of the partial waves can be demonstrated by using the asymptotic expansion of the Bessel function towards infinity, and by choose the initial condition
the very same partial wave. In this case the coefficients $T_{m,n}$ are proportional to the expressions
$$
T_{m,n}\sim \frac{e^{2 \pi i (m_0 -m)}-1}{m_0 -m}\biggl[C_1 \frac{e^{i \frac{\xi_{m,n}-a_0 L}{L}}}{\xi_{m,n}-a_0 L} +C_2 \frac{\hbox{erf} [(1+i) \sqrt{\xi_{m,n}-a_0 L}]}{(\xi_{m,n}-a_0 L)^{3/2}}  \biggr],
$$
where $C_{1,2}$ are normalization constants and erf is the error function. The first factor in the above evaluation of $T_{m,n}$ has its Fourier
transform the unit step function with support $|m_0 -m| \le 2 \pi$ so it is only
relevant in the IFF series if the summation index $m$ is in a $\pm 2 \pi$ neighborhood of $m_0$. The first term in the parenthesis is subjected to the same behavior. Its contribution is relevant only if $|\xi_{m,n}-a_0 L|\le 1$ which means only those $(m, n)$ pairs (actually the $(m_0 , n)$ pairs) who fulfill this inequality are dominant. Finally, the last term in the parenthesis has a $\delta-$Dirac behavior. In a neighborhood of $x \sim 0$ it behaves like $x^{-1/2}$ and it has an exponential decay towards infinity, as one can see from its Taylor series around any point $x_0 \neq 0$. These arguments demonstrate that out of
the whole series from Eq. (\ref{eq25}) only $m_0$ term and other  very few $n$ terms fulfilling the condition $|\xi_{m_0 ,n}-a_0 L|$ are relevant. It means that towards  large time values  practically the solution reconstructs the initial condition partial wave.

\subsection{Time evolution of the asymptotic solution}
\label{subsecTimeAsympt}

%\begin{figure}
%\includegraphics[scale=1]{f1.eps}
%\centering
%\caption{Archimedean spiral with one arm, and decaying profile as $1/r$.}
%\label{fig1}
%\end{figure}

If we consider the limit $L \rightarrow \infty$, which is a natural limit given the large scale of observed circular ice formations, \cite{1,2}, the solution Eq. (\ref{eq25}) has the asymptotic expression
\begin{equation}\label{eq34}
\Phi^1 \sim \frac{2 }{L^2 } \sqrt{\frac{2}{\pi }}\sum_{m=-\infty}^{\infty}\sum_{n=1}^{\infty} \frac{ c_{m} \biggl(  \mathcal{F}_{c} \biggl[ \sqrt{s} \Phi_{0}(s,\phi) \biggr]_{\frac{\xi_{m,n}}{L}} \biggr) }{J_{m}^{2} ( \xi_{m,n} ) \ r} \cos \biggl[\frac{\xi_{m,n}(r+ct)}{L}+m \theta \biggr] 
\end{equation}
where we used the asymptotic representation of the Bessel functions $J_m (x)\rightarrow \sqrt{2/(\pi x)} \cos(x-m \pi /2-\pi/4)$, and where $c_m$ represent the complex Fourier series coefficients of the function in the parenthesis (as a function of $\phi$), and $\mathcal{F}_c$ is the Fourier cosine  of the argument as a function of $s$ evaluated at $\xi_{m,n} / L$. That is
$$
c_m ( \mathcal{F}_{c} [\sqrt{s}\Phi_0(s,\phi)])=c_m(\hat{[\sqrt{s}\Phi_0 ]}(\zeta,\phi)) \rightarrow c_m(\hat{[\sqrt{s}\Phi_0 ]}(\xi_{n,m}/L,\phi))=c_m (n,m).
$$
This result is a very good tool to classify the stability of different initial spiral configurations. According to the definition  of a Archimedean spiral pattern in our physical system in the beginning of section 4, the initial distribution of ice fragments has the general form
$$
\Phi_{0}=g(r) P(a r+m_0 \theta),
$$
with $g(r)$ a decreasing function describing the diffusion of the spiral towards the boundaries into the sea, and $P$ o periodic function which governs the geometry of the spiral and its rotation. To be able to model we will consider for $g$ various powers of $1/r^\alpha, \alpha >0$ and for $P$ a cosine. From the asymptotic expansion formula Eq. (\ref{eq34}) we  calculate the distribution of the coefficients of the series. There are only two situations of stable spirals. 
One is if $\alpha =0$, that is uniform distributed spiral. In this pattern the ice fragments do not decrease their density, or the width of spiral arms,   larger $r$. The other case of stability is when $\alpha=1$ and the spiral radially decays  as $1/r$. In the first case the Fourier cosine transform in Eq. (\ref{eq34}) is applied to $r$ times trigonometric function. In the second case the transform is applied just to the trigonometric function. In both these situations the Fourier cosine transform result is the derivative of a $\delta-$Dirac distribution, or the $\delta-$Dirac distribution times constants. The result of the transform is a constant times $\delta (\xi_{m,n}-aL)$. This result restrict the sum over $n$ to only one term $n_0$, namely the solution of the transcendental equation $\xi_{m,n_0}=aL$, for each $m$. The next step is the calculation of complex Fourier coefficient $c_m$, but this turns to be trivial, since by orthogonality, the only contributing term filtered from the series over $m$ is $m_0$. This means that the time dependent solution $\Phi^1$ maintains the pattern of the initial condition, and stabilizes the Archimedean spirals of this types.
It results that out of the double series in Eq. (\ref{eq34}), only the term with $m=m_0$ and $n=n_0$ are contributing.
%\begin{figure}
%\includegraphics[scale=1]{f4.eps}
%\centering
%\caption{Plot of the linear solution $\Phi^1 (r,\theta,t)$ generated by %initial condition of a partial wave $\Phi_0 =\Phi^{1}_{m=1,n=1}$ at $t=L/c$. %The initial Archimedean spiral with one arm and decaying profile as %$1/\sqrt{r}$ tends to slightly deform, yet still keeping a similar shape.}
%\label{fig3new}
%\end{figure}

\begin{figure}
\includegraphics[scale=.25]{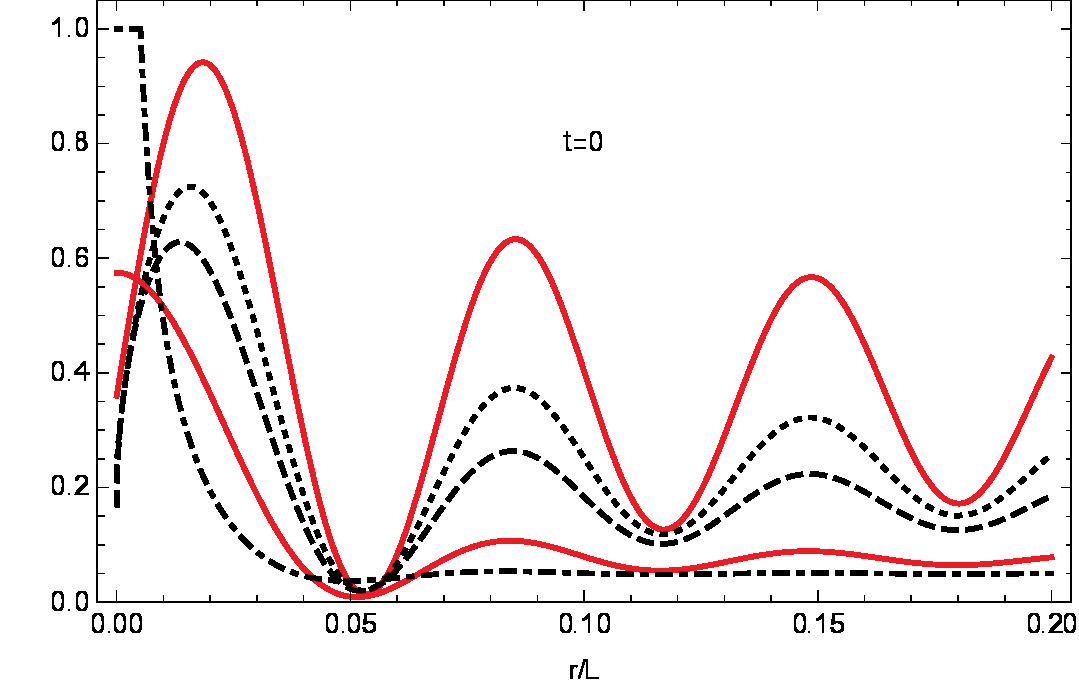}
\centering
\caption{Cross section in the $\theta=0$ vertical plane for the functions $\Phi^{1}(r,0,0)$ as solutions of the initial condition problems $\Phi_{0}=J_{1}(3 r/L)$ in solid red,  $r^{-0.3}J_{1}(3 r/L)$ dotted curve, $r^{-1/2}J_{1}(3 r/L)$ dashed curve, $r^{-1}J_{1}(3 r/L)$ solid red,  and $r^{-2}J_{1}(3 r/L)$ dotted-dashed curve. All curves represent Archimedean spiral with the same pitch but various decay laws with $r$.}
\label{fig4new}
\end{figure}

\begin{figure}
\includegraphics[scale=.25]{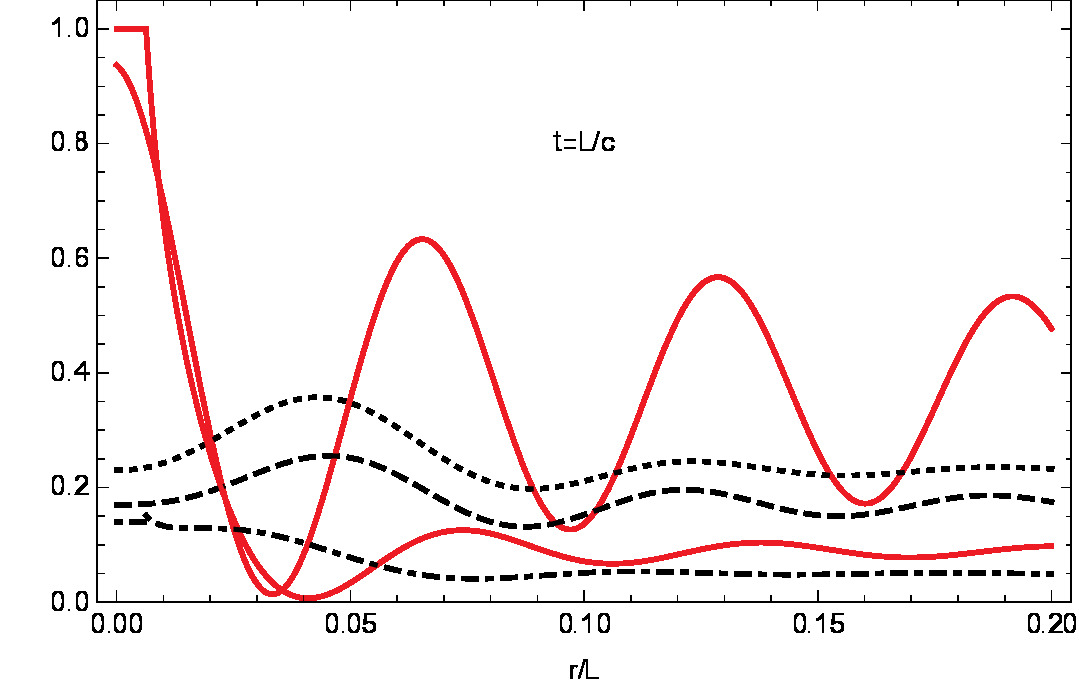}
\centering
\caption{The same cross section and the same initial conditions as in Fig. \ref{fig4new}, here the time evolution at the moment $t=L/c$. The black curves disintegrate the initial spiral profile, while the red curves are just translated which represents a $\theta$ rotation of the whole spiral.}
\label{fig5new}
\end{figure}

%\begin{figure}
%\includegraphics[scale=1]{f3.eps}
%\centering
%\caption{Archimedean spiral with one arm, high value of the pitch (large %value for $m$) and decaying profile as $1/r^2$}
%\label{fig6new}
%\end{figure}

More importantly, this finding means that the only Archimedean spirals of the types 
$$ 
\Phi_0 \sim \cos (a r+m_0 \theta), \ \ \ \Phi_0 \sim \frac{\cos(a r+m_0 \theta)}{r},
$$
which fulfill  the special relation between their pitch and radial extension $L$ (size of the swirl)), falling in one of the zeroes of the Bessel functions, and given by
$$
\hbox{pitch}=\frac{2 \pi m_0 L}{\xi_{m_0,n_0}}, \ \ n_0 \ \hbox{solution of} \ \xi_{m_0,n_0}=aL,
$$
are linearly stable.

Other types of Archimedean spirals, of the class we investigate here, may have a different type of decay law of the form $1/r^\alpha$ with $\alpha \in (0,1)$, which includes $1/\sqrt{r}$. These radial dependence generate Fourier cosine transform with rational functions dependence of the variable $\zeta$, with singularity point also at $\xi_{m_0,n_0}=aL$. The Fourier cosine spectrum is concentrated in this singularity as in the previous cases, but there are also tails which mix the initial condition in time with  other  partial waves. The initial spirals are weakly stable, and tend to slowly disperse in time, without actually completely erasing the spiral pattern.  Among these situations we have  the decay type of the asymptotic expression for the Bessel functions as $1/\sqrt{r}$. This result explains why single partial waves, involving one Bessel function, cannot form sharp stable spirals and tend to mix modes and deform the pattern. 

An example showing five types of decay law with $r$ are presented in Fig. \ref{fig4new} at initial moment of time (Archimedean spirals of the same pitch), and in Fig. \ref{fig5new} at a later moment in time, when the spiral with too slow or too steep decay law disintegrated. Finally,  Archimedean spirals with steeper decay with a dependence of the form $1/r^\alpha$ with $\alpha   >1$ generate a wide band Fourier cosine spectrum, and some higher power generate even divergent integrals. All these steeper decay laws  are unstable and disperse quickly in time. 

The results presented in this subsection demonstrate that, at least in the linear
approximation of the dynamical equations and in the inviscid case, only some selected Archimedean spiral patterns tend to be stable for large scales of space and time.

\subsection{Time Evolution of Archimedean spirals}
\label{subsecTimeArchm}

In this subsection we choose for initial condition more general form of Archimedean spirals, namely
$$
\Phi_0 = h(r) g(ar+m_0 \theta),
$$
where the function $h$ must decrease with $r$ and $g$ is periodic. We plug this initial condition in the general solution (not the asymptotic approximation) Eq. (\ref{eq25}). The double series coefficients contain the form
$$
T_{m,n} \sim \int_{0}^{2 \pi} \int_{0}^{L} s h(s) g(as+m_0  \Phi) J_{m}\biggl( \frac{\xi_{m,n}}{L}s \biggr) e^{i m \phi} ds d\phi.
$$
We make a change of variable $q=as+m_0 \phi$, use the fact that $g$ is $2\pi$-periodic to translate the limits of integration with respect to $q$, and the above  integral becomes
$$
T_{m,n} \sim \int_{0}^{2 \pi m_0 \phi} g(q) e^{-i\frac{q m}{m_0}}\biggl( \int_{0}^{L} s h(s) J_{m}\biggl( \frac{\xi_{m,n}}{L}s \biggr) e^{i \frac{m s a}{m_0}} ds \biggr) dq
$$
We can expand the function $h$ in its Taylor series around a point between $0$ and $L$, make the same supposition that $L$ is large enough to consider the upper limit of the integrals in $s$ as infinity,  and consequently the inside integral is just a sum of inverse Fourier transforms of a power law  time a Bessel function. These inverse Fourier transforms output as function of $\zeta$ the product between a finite support step function, and a polynomial. The integral over $\phi$ also reduces to a $\delta-$Kronecker symbol. We have
$$
T_{m,n} \sim \pi H \biggl( \frac{\xi_{m,n}}{L}-a\frac{m  }{m_0} \biggr)  \mathcal{P}_{o(m)}\hat{g}(1)\biggl( a\frac{m}{m_0} \biggr) \delta_{m,m_0},
$$
where $H$ is the Heaviside distribution, $\delta$ is the Kronecker symbol, and $\mathcal{P}_{o(m)}$ is a polynomial of order $o(m)$ which is $m$ if $m$ is even, and $m+1$ is $m$ is odd. The hat means the Fourier transform. This term filters out of the double series the terms with $m=m_0$ and $n > n_0$ where $n_0$ is the solution of the transcendental equation $\xi_{m_0 ,n}>a L$, exactly as in the result obtained in subsection 4.2. The solutions Eq. (\ref{eq25}) for the initial condition $\Phi^{1}(r, \theta, 0) = h(r) g(ar+m_0 \theta)$ becomes
$$
\Phi^1 =2 \pi \mathcal{P}_{o(m_0)}(a)\hat{g}(1) e^{i m_0 \theta}\sum_{n=n_0}^{\infty}  \frac{J_{m_0}\biggl( \frac{\xi_{m_0,n}}{L}r \biggr)}{L^2 J_{m_0 +1}^2(\xi_{m_0 ,n})} e^{i \frac{c \xi_{m_0,n} }{L}t}.
$$
We would like to evaluate the geometric parameters of the Archimedean spiral in a real situation. In all previous subsections we noticed that the solution series begins with a first term for $n=n_0$. Since the series must be convergent, we know the first term is the dominant one. Another noticed fact is that the real case spiral has one, maximum two arms, so $m_0=1$ or $2$. The core of the spiral, which usually is ice so $\Phi(0) \sim 1$ can be considered to extend radially from origin up to the first zero of the Bessel function involved in the solution. We can obtain the radius of the core from the equation $J_{m_0} (\xi_{m_0, n_0} r_c /L) \simeq J_{m_0}(\xi_{m_0, 1})$ so we have  $r_c \simeq \xi_{m_0,1} L/\xi_{m_0, n_0}$. We can take the radius of the core to be at half way between origin and first zero, so $r_c \simeq \xi_{m_0,1}L/(2 \xi_{m_0,n_0})$. Also from the dominant term we can roughly estimate the radius of the swirl. Let us assume that the swirl ends when IFF becomes $\epsilon \Phi$ evaluated at the core. From here we obtain $L \simeq \xi_{m_0, n_0}/(2 \epsilon^2)$. The pitch can also be evaluated as pitch$=2 \pi m_0 / \xi_{m_0,n_0}\simeq (2 \pi \xi_{m_0,n_0}) L$. The 
documentation shows spirals with $5-10$ turns, so $L/$pitch $\simeq 5-10$. If we choose $\epsilon=0.05$ we have $\xi_{m_0,n_0} \simeq 2 \pi m_0 L/$pitch which ranges $\xi_{m_0,n_0}$ between $30$ and $120$. It results for $m_0=1$ or $2$ the possible range $n_0=10-38$. From here we can estimate $L\simeq 5 10^3 \xi_{m_0,n_0}\simeq 50-230$Km with core $r_c \simeq (\xi_{m_0,1}/\xi_{m_0,n_0}) L\simeq 2-40 $Km, of course function of the radius of the swirl.

%%%%%%%%%%%%%%%%%%%%%%%%%%%%%%%%%%%%%%%%%%%%%%%%%%%%%%%%%%%%%%%%%%%%%%%%%%%%%%
%%%%%%%%%%%%%%%%%%%%%%%%%%%%%%%%%%%%%%%%%%%%%%%%%%%%%%%%%%%%%%%%%%%%%%%%%%%%%%
\section{Limiting nonlinear modes}

There are three limiting types of axially symmetric flows (that is neglecting plane density waves at the surface) for the ice and water mixture: the \textit{radial mode} when we can neglect azimuthal flow and rotations, and the whole system performs radial oscillations of compression and dilation; the \textit{rotational mode} in which the system is in rotational flow and we neglect radial components, and the \textit{spiral mode} when the flow is concentrated mainly along a certain spiral direction field.

\subsection{Radial limiting mode}

In this situation we neglect the azimuthal velocity $v$ terms in Eqs. (\ref{eq9}, \ref{eq10}) and, with the notations $\Psi=1-a \Phi$, $b=(P_{w} / \rho_{w}-gT_0 )/a$, we have
\begin{equation}\label{eq9radia}
\Psi_t+ (u\Psi)_r +\frac{u \Psi}{r}=0,
\end{equation}
\begin{equation}\label{eq10radia}
(\Psi u)_t+ u (\Psi u)_r -b\Psi_r=0,
\end{equation}
while the last Eq. (\ref{eq11}) reduces to $\Psi_{\theta}=0$. It is the most natural in this case to assume that a the fields ($\Psi, u$) have radial symmetry. With the notation $\bar{u}=\Psi u$, and multiplying the second equation with $\Psi$ we have 
\begin{equation}\label{eq9radia1}
\Psi_t+ \bar{u}_r +\frac{\bar{u}}{r}=0,
\end{equation}
\begin{equation}\label{eq10radia1}
\Psi \bar{u}_t+\bar{u} \bar{u}_r -b \Psi \Psi_r =0,
\end{equation}
equations representing conservation laws for surface mass density and mass flow \begin{equation}\label{eq9radia2}
\iint_{\mathbf{R}^2} \Psi dA =\hbox{const.}, \ \ \iint_{\mathbf{R}^2} \Psi \bar{u} dA =\hbox{const.},
\end{equation}
respectively, under the assumption that the radial velocity decreases towards infinity more rapid that $1/r$. From a brief nonlinear dispersion relation analysis, we note that the system of equations is scale invariant to the magnitude of $\Psi$,  the mean value of the radial velocity  ranges   $<u> \sim \sqrt{b} =\sqrt{(P_w-\rho_w g T_0)/(\rho_w -\rho_i)}$ and the radial oscillation frequency ranges $\omega/(2 \pi) \sim <u>/R \simeq \sqrt{b}/R$, where $R$ is the radius of the ice-water mixture. 

The radial mode system of equations in the form Eq. (\ref{eq9radia1}, \ref{eq10radia1}) is separable, and the resulting PDE for $\bar{u}(r,t)$ has a very busy expression
$$
Y \bar{u}_t - b Y_{r} Y + \Psi_{1}(t) \bar{u}_t - b \Psi_{1}(t) Y_{r} +\bar{u} \bar{u}_t =0,
$$
where $\Psi_{1}$ is an arbitrary function of $t$ only, and $Y$ is a nonlinear expression in $\bar{u}$ and its derivatives
$$
Y=\int \biggl[ 2 b r \bar{u}\bar{u}_t+b r^2 \bar{u}_t \bar{u}_r-b r^2 \bar{u}\bar{u}_{r,t} \pm [(b r^2 \bar{u}\bar{u}_{r,t}-2br \bar{u}\bar{u}_t -b r^2 \bar{u}_r \bar{u}_t )^2
$$
$$
-4(b^2 r \bar{u}+b^2 r^2 \bar{u}_r )(r \bar{u}\bar{u}_{t}^{2}-b \bar{u}^{2} \bar{u}_r +r^2 \bar{u}\bar{u}_{tt}\bar{u}_r +br \bar{u}\bar{u}_{r}^{2}-r^2 \bar{u}\bar{u}_t \bar{u}_{r,t}+b r^2 \bar{u}\bar{u}_r \bar{u}_{rr})]^{\frac{1}{2}} \biggr] 
$$
$$
\times (2 b^2 r \bar{u}+2b^2 r^2 \bar{u}_r)^{-1} dr
$$

Nevertheless, since the first equation in the system Eq. (\ref{eq9radia1}) is originally linear, we apply  perturbative methods to the above system depending only on the two unknown functions $\Psi, \bar{u}$, namely $\Psi \simeq \Psi_0 +\Psi_1, \bar{u} \simeq \bar{u}_0 +\bar{u}_1 , \bar{u}_r \simeq \bar{u}_{0r} +\bar{u}_{1,r}$. We linearize of Eq. (\ref{eq10radia1}), differentiate it to $t$, and couple it with Eq. (\ref{eq9radia1}) differentiated to $r$, and we obtain in the zero order $\bar{u}_0 =\sqrt{b} \Psi_0$, and in the first order
\begin{equation}\label{eq10lineariz}
\frac{1}{b}\bar{u}_{1,tt}+\bar{u}_{1,rr}+\frac{\bar{u}_{1,r}}{r}-\frac{1}{r^2}\bar{u}_{1}+\frac{\bar{u}_0 \bar{u}_{0,r}}{b \Psi_0}\bar{u}_{1,rt}+\frac{\bar{u}_{0,r}}{b \Psi_0} \bar{u}_{1,t}+\frac{\bar{u}_{0} \bar{u}_{0,r}}{b \Psi_{0}^{2}}\biggl( \bar{u}_{1,r}+ \frac{\bar{u}_{1}}{r}\biggr)=0.
\end{equation}
By considering the Fourier decomposition and substituting $\partial_{t} =i \omega $Id, we obtain from Eq. (\ref{eq10lineariz}) a generalized hypergeometric differential equation. The solution reads
\begin{equation}\label{eqhyperlinu1}
\bar{u} =\bar{u}_0+r \exp \biggl(-\frac{k_1 r}{2 \Psi_{0}^{2}} \biggr) \biggl[ C_1 U\biggl(-k_2,3,\frac{r k_1}{\Psi_{0}^{2}} \biggr) +C_2 L_{k_2}^{2} \biggl( \frac{r k_1}{\Psi_{0}^{2}} \biggr) \biggr] e^{i \omega t},
\end{equation}
where $U$ is the confluent hypergeometric function, $L$ is the generalized Laguerre function, $C_{1,2}$ ar constants of integration, and we have  the parameters
$$
k_1=\sqrt{\omega \Psi_{0}} \sqrt{4b \omega \Psi_{0}^{3}+\bar{u}_{0}^{2} \bar{u}_{0r}^{2} (1+2i-\Psi_0 \omega) -4i\bar{u}_{0r}\Psi_{0}^{2}},
$$
$$
k_2=\frac{\bar{u}_{0} \bar{u}_{0r}}{2 k_1}(1-i \omega \Psi_{0})-\frac{3}{2},
$$
$$
k_{3}=\frac{\bar{u}_{0} \bar{u}_{or}}{b}(1-i \omega \Psi_{0}).
$$
To simplify further this solution, we can neglect the contribution of the term $\bar{u}_{1} \bar{u}_{1,r}$ from Eq. (\ref{eq10radia1}) and the corresponding simplified solution is expressible in term of the modified Bessel functions
$$
\bar{u}=\bar{u}_0 + \biggl[ C_1 I_{1}\biggl( \frac{r \omega}{\sqrt{b}} \biggr)+C_2 K_{1}\biggl( \frac{r \omega}{\sqrt{b}} \biggr) \biggr] e^{i \omega t}.
$$
The calculations are complete by implementing the above solutions for $\bar{u}$ in Eq. (\ref{eq9radia1}), integrating with respect to time to obtain the IFF field $\Psi$
$$
\Phi(r,t)=\mathit{Re}\biggl\{ -\frac{i}{\omega} \biggl( \bar{u}_{r}+\frac{\bar{u}}{r} \biggr) e^{i \omega t} \biggr\},
$$
and substituting this $\Psi$ solution $u=\bar{u} / \Psi$ to finally obtain the radial velocity. 

From the linear stability analysis of the system Eqs. (\ref{eq9radia1}, \ref{eq10radia1}) we obtain that always one of the linear equilibrium solutions is  asymptotically stable, and the second always unstable. This fact however, does not guarantee the nonlinear stability of the system all together. For a large range of parameters the IFF solution is always around unit at origin and decreases parabolic towards large distances, while the radial velocity is zero at origin and increases almost linear towards the periphery of the ice disk, and both oscillate coherently in the radial direction. The resulting radial mode consists in very slowly periodic radial compression and dilation with the rate of change increasing with distance to center. Typical value are $\Phi (r=30 \hbox{Km})\sim 0.85, u \simeq 100 \div 2000 $ m/day, and $2 \pi / \omega \simeq 6\div 40$ hours.

\subsection{Rotational limiting mode}

In this limiting mode we neglect radial velocity $u\simeq 0$ and Eqs. (\ref{eq9}) amd (\ref{eq11}) acquire the simplified form
\begin{equation}\label{eq33333}
\Psi_t+\frac{(\Psi v)_{\theta}}{r}=0,
\end{equation}
\begin{equation}\label{eq33333x}
(\Psi v)_t +\frac{\Psi v v_{\theta}}{r}+\frac{\Psi_{\theta} v^2}{r}=-\frac{b}{ar}\Psi_{\theta},
\end{equation}
while Eq. (\ref{eq10}) can be integrated into $v=\pm\sqrt{b r \Psi_r /(a \Psi)}$. Such solution exists only if $\Psi_r \ge 0, \Phi_r \le 0$ meaning that ice always accumulates at the center of the rotating pattern. By differentiating the equations above to $\theta, t$ respectively we obtain a nonlinear equation for $\Omega =\ln (\Psi)$
\begin{equation}\label{eq33333y}
r\frac{a}{b}(\Omega_{tt}+\Omega_{t}^{2})=\frac{1}{r}(\Omega_{\theta \theta}+\Omega_{\theta}^{2})+\Omega_{r \theta \theta}+\frac{3}{2}\Omega_{\theta} \Omega_{r \theta}+\Omega_r \Omega_{\theta \theta}+\Omega_{\theta}\Omega_{r \theta}+\Omega_r \Omega_{\theta}^{2}. 
\end{equation}
This equation is similar to the Nonlinear Schr\"{o}dinger equation, but it contains several other terms. A simple solution for this equation is the rigid rotation of the whole pattern, $v=\omega_0 r$, which is also a solution of  Eq. (\ref{eq33333}). In this case Eq. (\ref{eq33333x}) can be easily integrated given the observation that $b \ll a r^2 \omega_{0}^{2}$ and finally generates 
$$
\Phi=\frac{1}{a}\biggl( 1-P(\theta - \omega_0 t)e^{\frac{a \omega_{0}^{2} r^2}{2b}}\biggr),
$$ 
where $P$ is an arbitrary function. This rotational pattern and $r$ dependence confirms the accumulation of ice towards the center. While looking for other possible solution we notice that in the long range, the asymptotic form of Eq. (\ref{eq33333y}) for $r \rightarrow \infty$ would have a blow-up solution $\Psi \sim t$. The equilibrium solutions $\partial_{t} =0$ of the system Eqs. (\ref{eq33333}, \ref{eq33333x}) represent un-physical situation. Indeed it results $(\ln \Omega_{\theta})_{r}+2 \Omega_r+r=0$ and hence the corresponding solutions $\Psi \sim \sqrt{\theta} \exp (-r^2)$ are not periodic in $\theta$, hence are strongly unstable.
%%%%%%%%%%%%%%%%%%%%%%%%%%%%%%%%%%%%%%%%%%%%%%%%%%%%%%%%%%%%%%%%%%%%%%%%%%%%%%
%%%%%%%%%%%%%%%%%%%%%%%%%%%%%%%%%%%%%%%%%%%%%%%%%%%%%%%%%%%%%%%%%%%%%%%%%%%%%%

\subsection{Spiral limiting mode}

If the flow follows the geometry of a family of time-independent logarithmic spirals of equations $R(\theta)=R_0 \exp{g_0 \theta}$ we have the simplifying relation $u=R_{\theta} v/R=g_0 v$. With this substitution for $u$, by dividing Eq. (\ref{eq10b}) by $g_0$ and subtract it from Eq. (\ref{eq11b}) we obtain
\begin{equation}\label{eqmm4354}
\frac{b}{ar}\Psi_{\theta}+\frac{g_0}{r} \Psi v^2 -\frac{b}{a g_0}\Psi_r +\frac{1}{r g_0}\Psi v^2 =0,
\end{equation}
or
\begin{equation}\label{eqmm4354b}
\frac{\Psi_r}{\Psi}=\frac{v^2 a g_0}{rb} \biggl( \frac{1}{g_0}+g_0\biggr)+\frac{g_0}{r}\frac{\Psi_{\theta}}{\Psi}.
\end{equation}
If we assume the simplest periodic dependence of $\Psi$ by the azimuthal angle such that $\partial_{\theta}\Psi=i m \Psi$ we have after one quadrature
\begin{equation}\label{eq49bv28p94tpv}
\Psi=\exp{\biggl[ \frac{a }{b}\biggl(g_{0}^{2}+g_0 \biggr) \int^{r} \frac{v^2(\xi,\theta, t)}{\xi}d\xi\biggr]} \exp{[i(m g_0 \ln r+ m \theta)]} 
\end{equation}
This is an integral representation of the solution, but it reveals a stable dependence on the logarithmic spiral argument, modulated with an exponential depending only on the tangent velocity. Most importantly, since the multiplicity $m$ factors out, only one-arm logarithmic spirals are possible 

From the remaining equations Eq. (\ref{eq9b}) and Eq. (\ref{eq10b}), also under the substitution $u=g_0 v$, and expressed as time Fourier transform ($\partial_{t}=i \omega$)  we obtain
\begin{equation}\label{eqnqnvwtoppw}
i \omega \Psi+g_0 (\Psi v)_r +\frac{2 i m\Psi v}{r}+\frac{g_0}{r}\Psi v=0,
\end{equation}
\begin{equation}\label{eqnqnvwtoppwb}
2 i \omega \Psi v+g_0 v (\Psi v)_r +\frac{2 i m\Psi v^2}{r}+\frac{g_0}{r}\Psi v^2 +\frac{i m b}{ar}\Psi=0.
\end{equation}
Once we proved that the ice pattern follows a stable logarithmic spiral, from here the procedure to obtain the velocity field is direct. By replacing $(\Psi v)_r$ in Eq. (\ref{eqnqnvwtoppwb}) from its expression in Eq. (\ref{eqnqnvwtoppw}), we obtain an algebraic equation  in the functions $\Psi$ and $v$
\begin{eqnarray}\label{eqnvnfjf}
\frac{imb \Psi}{ar}+\Psi v \biggl[ 2i \omega +\frac{2 i m v}{r}+\frac{v g_0}{r}\biggr] -v g_0 \biggl[ \frac{i \omega \Psi}{g_0}+\frac{\Psi v}{r g_0}(2 i m+g_0 )\biggr]=0
\end{eqnarray}
We solve this algebraic equation for $v=v(\Psi, r)$ and substitute this solution in Eq. (\ref{eq49bv28p94tpv}) transforming this equation into an integro-differential equation for $\Psi$ only, and with $\Psi$ obtained find $v$ and $u$.

As a final observation in this section, we mention that  we can find a very stable and simple form for the logarithmic spiral solution in a particular equilibrium situation. If we assume $b \ll v^2$ in Eqs. (\ref{eq9b}, \ref{eq11b}), both these equations become identical to
$$
(\Psi v)_r =-\frac{\Psi v}{r g_0} (2im+g_0 )-\frac{i \omega \Psi v}{g_0},
$$
which generates a solution in the form
$$
\Psi v= \frac{C}{r}e^{m i (-2 \ln r /g_0 +\theta)}.
$$
This is again the structure of a stable one-arm logarithmic spiral. The condition for small $b= P_{w} / \rho_{w}-g T_0$ represent a case when the phase velocity of the density waves through the sea-ice pattern is way smaller than the group velocities, and actually means that the water pressure is balanced by the ice buoyancy, and there is no strong interaction and exchange of momentum between ice and water. It also means that the gradient of pressure terms in the Navier-Stokes equations are negligible, which means that logarithmic spiral patterns tend to form and stabilize when the sea-ice is rather isolated from currents or winds.

%%%%%%%%%%%%%%%%%%%%%%%%%%%%%%%%%%%%%%%%%%%%%%%%%%%%%%%%%%%%%%%%%%%%%%%%%%%%%%
%%%%%%%%%%%%%%%%%%%%%%%%%%%%%%%%%%%%%%%%%%%%%%%%%%%%%%%%%%%%%%%%%%%%%%%%%%%%%%

\section{The nonlinear equations}
\label{secfullnl}

Eq. (\ref{eq9b}) has exact time-independent solutions generating logarithmic spiral patterns
\begin{equation}\label{eq.solutnl}
\Psi=f(r) e^{i [g_0 \ln (r/r_{0}) +m \theta ]}, \ \ u(r)=-\frac{m q_0 r_0}{g_0 r f(r)}, \ \ v(r)=-\frac{g_0}{m} u(r),
\end{equation}
where 
$$
f(r)=\frac{r_0 \sqrt{a q_{0}^{2} (g_{0}^{2}+m^2)+2 b g_{0}^{2} f_0 r^2}}{\sqrt{b} g_0 r},
$$
and $g_0 , r_0, q_0, f_0$ are arbitrary real parameters and where the value of $a$ must be adjusted such that $\Phi \in [0,1]$.  The IFF distribution continuously decreases from center towards its asymptotic value  
$$
\Phi_{\infty}=\frac{\rho_{w} (1-r_0 \sqrt{2 f_0})}{\rho_w - \rho_i},
$$
so the boundary conditions around the spiral determine the parameter $r_0 \sqrt{f_0}$. We can choose a minimal radius $r_{min}$ above which the model validates, and since in all observations the center of the spiral appears to me solid ice we can determine another condition $\Phi (r\simeq 0)\sim 1 \sim r_0 q_0 \sqrt{a (g_{0}^{2}+m^2)}/(g_0 r_{min} \sqrt{b})$. The ratio between the radial and azimuthal velocities is a constant determining uniform flow along spiral curve coordinates,  $r=r_0 \exp (m \theta/g_0)$. The velocity is also independent of time and azimuthal angle. 

Solution Eq. (\ref{eq.solutnl}) is almost an exact solution for the momentum conservation equations, Eqs. (\ref{eq10b}, \ref{eq11b}), within an approximation of order $\mathcal{O} (\sqrt{b}/r^2)$ which is about $10^{-9}$ times smaller than the order of the $\Psi$ solution, for example, within a large range of parameters values.  The parameter $r_0$ was introduced to allow dimension free argument in the logarithm, but also controls the smallness of the defect of this solutions in Eqs. (\ref{eq10b}, \ref{eq11b}).

With the notations $b=(P_{w}/\rho_{w}-g T_{0})/a, \Psi=1-a\Phi$, and similarity substitutions $\xi=\ln r, s=t/r = t \ \hbox{exp}(-\xi)$, Eqs. (\ref{eq9}-\ref{eq11}) can be written in a coefficients-free  symmetric form 
\begin{equation}\label{eq.finals41}
\Psi_s+\Psi u +(\Psi u)_{\xi} +(\Psi v)_{\theta}=0
\end{equation}
\begin{equation}\label{eq.finals4}
(\Psi u)_s+v (\Psi u)_{\theta}+u (\Psi u)_{\xi}- \Psi v^2+b\Psi_{\xi}=0,
\end{equation}
\begin{equation}\label{eq.finals42}
(\Psi v)_s+ v (\Psi v)_{\theta}+u (\Psi v)_{\xi}+\Psi u v+b\Psi_{\theta}=0.
\end{equation}
The system can be written in an almost symplectic form by using the covariant gradient operator $\mathcal{L}=\partial_s +u \partial_{\xi}+ v \partial_{\theta}$ along the flow vector field, and define the gradient $\nabla=(-b \partial_{\xi}, -b \partial_{\theta})$ \cite{book2} 
\begin{equation}\label{eq.sys.operat}
\begin{pmatrix} 
\mathcal{L}\Psi & v \Psi \\
-v \Psi & \mathcal{L}\Psi 
\end{pmatrix}
\begin{pmatrix} 
u+v  \\
u-v   
\end{pmatrix}
=\nabla^{T}\Psi, 
\end{equation} 
Eq. (\ref{eq.finals41}) for mass conservation contains linear and quadratic nonlinear terms, while the last two equations (Eqs. (\ref{eq.finals4}, \ref{eq.finals42}) for momentum conservation) involve also cubic nonlinearity. We follow this idea to identify solutions for the full nonlinear system.

\subsection{Spiral patterns from the full nonlinear system}

By differentiating Eqs. (\ref{eq.finals4}, \ref{eq.finals42}) 
with respect to $\xi$, and $\theta$, respectively, by substituting the first term from each resulting equation into the $s-$differentiated Eq. (\ref{eq.finals41}) we obtain the nonlinear extension of Eq. (\ref{eq20})
\begin{equation}\label{eq.finals41bis}
\Psi_{ss} -b ( \Psi_{\xi}+ \Psi_{\xi \xi}+ \Psi_{\theta \theta})+\mathcal{N.L.}[\Psi,u,v]=0,
\end{equation}
where the velocities occur only in the higher orders terms, meaning that Eq. (\ref{eq.finals41}) describes mainly the geometry of the patterns through $\Psi$, while Eqs. (\ref{eq.finals4}, \ref{eq.finals42}) describe the dynamics of these patterns. Without any loss of generality we look for solutions appropriate for description of the  ice patterns with circular of spiral symmetry in the form $\Psi=f(\xi) \hbox{Exp}[i \chi_1 (\xi,\theta,s)], u=g(\xi) \hbox{Exp}[i \chi_2 (\xi,\theta,s)],v=h(\xi) \hbox{Exp}[i \chi_3 (\xi,\theta,s)]$ with $f,g,h,\chi_{1,2,3}$ real functions.

By implementing these forms in Eq. (\ref{eq.finals41bis}), we obtain from its real part
$$
[fg+(fg)_{\xi}] \cos \chi_2 -fh (\chi_{1, \theta}+\chi_{3,\theta}) \sin \chi_{3}-fg (\chi_{1, \xi}+\chi_{2,\xi}) \sin \chi_{2}=0.
$$
Under certain legitimate hypotheses we can use this equation to find the phase $\chi_1$ for the shape function $\Psi$. The remaining imaginary part of Eq. (\ref{eq.finals41}), and the real/imaginary parts of Eqs. (\ref{eq.finals4}, \ref{eq.finals42}) provide a system of 5 equations for the remaining functions $f, g, h, \chi_2 , \chi_3 , $ to be determined.

\begin{figure}
	\centering
	\includegraphics[width=7cm,height=7cm]{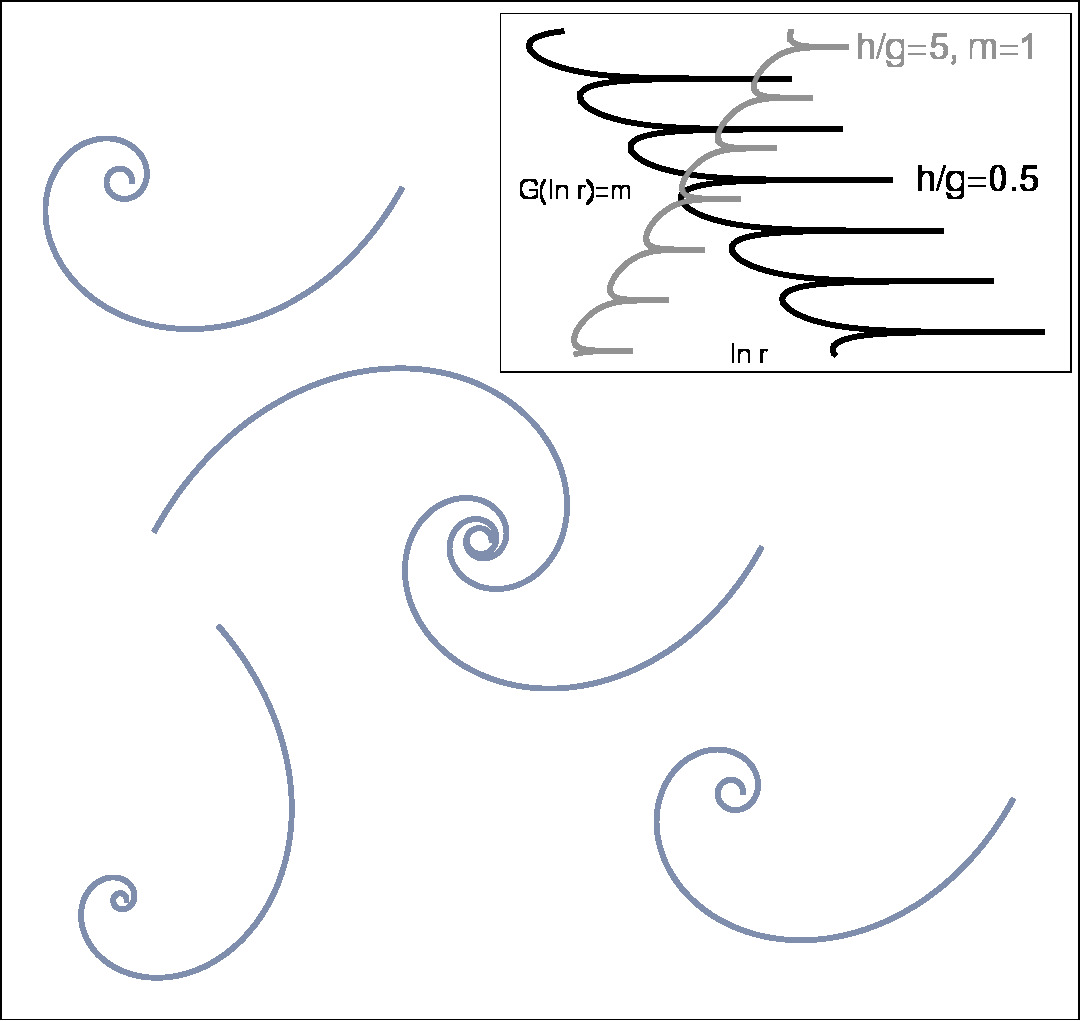} 
	\caption{Solution Eq. (\ref{eq.979586490}) for $m=1$ and two values for $a_0 = h/g=v/u$ for the nonlinear mass conservation equation describing coexisting logarithmic spiral patterns of ice in water. Corresponding spirals' equations $G (\ln r)=m \theta$ are plotted in the inset.}
	\label{figmany}
\end{figure}

\begin{figure}
	\centering
		\includegraphics[width=5cm,height=5cm]{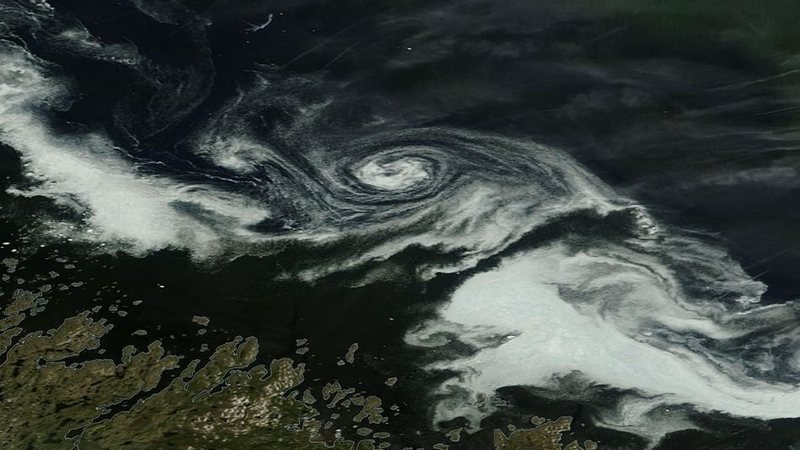} 
	\caption{Ice swirls in the Labrador current. Courtesy of NASA and \cite{moreswirls}.}
	\label{figswirl}
\end{figure}

If the amplitudes $f,g,h$ vary slowly with $\xi$, while the phases $\chi_{1,2,3}$ vary quickly with $\xi$ (see for example the models for spiral galaxies \cite{galaxy1,galaxy2})  Eq. (\ref{eq.finals41bis}) generates solutions with spiral symmetry in the IFF distribution at any instant of time, the shape of the spiral being given $\chi(\xi,\theta,s)=$const. In order to demonstrate the existence of such spiral solutions, and in accordance with to the above hypothesis on the rate of change of various functions with $\xi$, we can neglect the term $(fg)_{\xi}$ with respect to the $\chi$ derivatives. Moreover, we can approximate in this equation  $\chi_{1} \simeq \chi_{2} \simeq \chi_{3}$ denoted in the following $\chi$, because differences between the phase terms occur in the $4^{th}$ order. The above equation becomes
\begin{equation}\label{eq1298318923}
\cos \chi - 2 \biggl( \frac{h}{g}\chi_{\theta}+\chi_{\xi} \biggr) \sin \chi =0
\end{equation}
Because of the $\theta-$periodicity, the general spiral solution described by the equation $\chi(\xi, \theta,s)=$const. must be in the form $\chi=G(\xi ,s)-m \theta=$const., hence $\chi_{\theta}=m$, in polar coordinates, with $m$ integer describing the number of arms in the spiral. The spiral pattern is coherent in time only if the velocity field $(u,v)$ is directed along the tangent to the spiral pattern at any point, which involves the condition 
$$
\frac{u}{v}=-\frac{m}{r G_r},
$$
otherwise the spiral would radially dilate, shrink or disperse. For example, an Archimedean spiral is generated if $u/v \sim 1/r$, and a logarithmic spiral is generated if this ratio is constant. In the logarithmic spiral case, $h/g=a_0 =$ constant, Eq. (\ref{eq1298318923}) provides an implicit solution $\chi (\xi,\theta)=G(\xi)-m\theta$
\begin{equation}\label{eq.979586490}
\ln r=\xi=-\frac{4 a_0 m G(\xi) +2 \ln (\cos G(\xi) - 2 a_0 m \sin G(\xi))}{1+4 a_{0}^{2} m^2}+\hbox{const.}
\end{equation}
This solution generates a distribution of logarithmic spirals in the plane with centers along a straight line, Fig. \ref{figmany}, of equations $G(\ln r)=m \theta$. The constant value $a_0$ for the ratio $h/g$ is related to the  asymptotic values of $G \rightarrow G_{\infty}$ when $r \rightarrow \infty$, $a_0 = \cot^{-1} G_{\infty} /(2m)$ which describes the maximum angle of the spiral where the pattern stops $\theta \le G_{\infty}/m$.
Solution Eq. (\ref{eq.979586490}) also predicts the limiting size of the logarithmic spirals as depending on the ratio of the flow velocities
$$
\theta \le \theta_{max}=\frac{1}{m}\tan^{-1}\frac{u}{2vm}.
$$
Also, this solution can explain the occurrence of multiple separated ice swirls, as they were observed in the ocean, Figs. \ref{figswirl}, \cite{moreswirls}

In the following we present an algorithm to construct exact series solutions of the full nonlinear system Eqs. (\ref{eq.finals41}, \ref{eq.finals4}, \ref{eq.finals42}) by using the scaled series in Eq. (\ref{eq12}).  The procedure is rooted in the observation that in the linearized version, Eq. (\ref{eq20}), the system reduces to a Bessel equation in polar coordinates. When we implement the iterative algorithm to the full nonlinear system, we notice that the same Bessel differential operator occurs recurrently in all orders of smallness, property which induces a diagonal structure to the iteration algorithm, hence validating its efficiency.

Beginning from Eq. (\ref{eq.finals41bis}) we obtain the following expansion in the smallness parameter $\epsilon$
$$
\epsilon ( B[\Phi_{1}] +\mathcal{N}_{1}[u_{1},v_{1}] ) +\sum_{k \ge 2} \epsilon^k  ( B[\Phi_{k}]+\mathcal{N}_{1}[\Phi_{0},\dots,\Phi_{k-1},u_{1},v_{1},\dots,v_{2 k-1}] ) 
$$
\begin{equation}\label{eq.psipsi1}
+\delta \sum_{j\ge 1} \epsilon^{j} \mathcal{N}_{2}[\Phi_{0},\dots,\Phi_{j-1},u_{2},v_{2},\dots,v_{2j}]=0, 
\end{equation}
where the Bessel linear operator $B$ has the following action
\begin{equation}\label{eq.psipsi2}
B[\Phi_k]= \Phi_{k,ss} -c \biggl( \Phi_{k,\xi}+\Phi_{k,\xi \xi}+\Phi_{k,\theta \theta} \biggr),
\end{equation}
and $\mathcal{N}_{1,2}$ are specific nonlinear operators acting on the arguments placed in front of them. For example
$$
\mathcal{N}_{1}[u_{1},v_{1}]=(1-a \Phi_{0}) [v_{1}^1-v_{1,\theta}^2-u_{1,\xi}^2-2 u_{1,\theta}v_{1,\xi}-v_{1} (2 u_{1,\theta}+v_{1,\theta \theta}
$$
\begin{equation}\label{eq.psipsi3}
-2 v_{1,\xi}+u_{1,\xi \theta})
-u_{1} (v_{1,\theta}+u_{1,\xi}+v_{1,\xi \theta}+u_{1,\xi \xi})].
\end{equation}
Under the previously made scaling assumption $\mathcal{O}(\epsilon)=\mathcal{O}(\delta^2)$, the structure of Eq. (\ref{eq.psipsi1}) becomes self-consistent, and it can be solved by successive iterations following the Lindstedt-Poincar\'{e} method, \cite{odebook}, since for any $k\ge 2$ and $j \ge 1$, $\mathcal{O}(\epsilon^k)\neq \mathcal{O}(\delta \epsilon^j)$ are independent orders. Consequently, each term from each of the three main terms in Eq. (\ref{eq.psipsi1})  generate independent equations. Namely, each pair of terms with coefficient $\delta \epsilon^{k-1}$ and $\epsilon^{k}, k\ge 2$ generate system of two nonlinear complex differential equations in only 5  dependent variables $\{ \Phi_k, u_{2k-2},u_{2k-1},v_{2k-2}, v_{2k-1} \}$, since the other dependent variables occurring in these terms,  $\{ \Phi_0, \dots, \Phi_{k-1}, u_1, \dots, v_{2k-3} \}$, were solved in the previous pair of differential equations, corresponding to the coefficients $\delta \epsilon^{k-2}, \epsilon^{k-1}$.
The only exception is the order $\mathcal{O}(\epsilon)$ where the corresponding nonlinear equation has the form
\begin{equation}\label{eq.644347}
\Phi_{1,ss} -c \biggl( \Phi_{1,\xi}+\Phi_{1,\xi \xi}+\Phi_{1,\theta \theta} \biggr)+\mathcal{N}_{1}[u_{1},v_{1}]=0,
\end{equation}
and it can be solved for $\Phi_{1},u_{1}$ and $v_{1}$ by using the substitutions
\begin{equation}\label{eq.892465209}
\Phi_{1}=\alpha(\xi,\theta,s) e^{i \chi(\xi,\theta,s)}, \ \ u_{1}=\beta(\xi,\theta,s) e^{i \chi(\xi,\theta,s)}, \ \ v_{1}=\gamma(\xi,\theta,s) e^{i \chi(\xi,\theta,s)}, 
\end{equation}
where all the three physical fields assume the same phase dependence imposed by the cylindrical (and later on spiral) symmetry. From the real/imaginary, and even/odd separations (i.e. $\cos, \sin$) of the terms of this equation we obtain exactly four differential  equations for  four real functions $\alpha,\beta,\gamma,\chi$. In the next step, the resulting term of order $\epsilon \delta$ provides the differential equation from where one can integrate  $u_{2},v_{2}$ from $\Phi_{1}, u_1, v_1$. The next order, the term $\mathcal{O}(\epsilon^2)$ generates an equation for $\Phi_2$ function of the previously obtained  $\Phi_1, u_1, v_1$, the next term in order $\mathcal{O}(\epsilon^2 \delta)$ provides an equation for $u_3, v_3$ function of $\Phi_1, \Phi_2, u_1,\dots, v_2$, and so on.

An important observation is coming from the linearized Eq. (\ref{eq20}) for mass conservation, which depends only on the $\Phi$ variable. When we consider higher orders, and other nonlinear terms, no matter of the scaling or approximation procedure used, this equation will have the general form 
\begin{equation}\label{eq.sine.gord}
\Phi_{tt}-c^2 \triangle_{cyl} \Phi=\mathit{N}(\Phi, u, v, D \Phi, \dots),
\end{equation}
where $\triangle_{cyl}$ is the Laplacian in cylindrical coordinates,  $\mathit{N}$ is a nonlinear operator in the model dependent variables and their derivatives. Without any loss of generality we can request the IFF function to have the form
\begin{equation}\label{eq.substit}
\Phi=f(r,\theta, t) e^{i \chi (r, \theta, t)},
\end{equation}
\begin{figure}
	\centering
	\includegraphics[width=7cm,height=9cm]{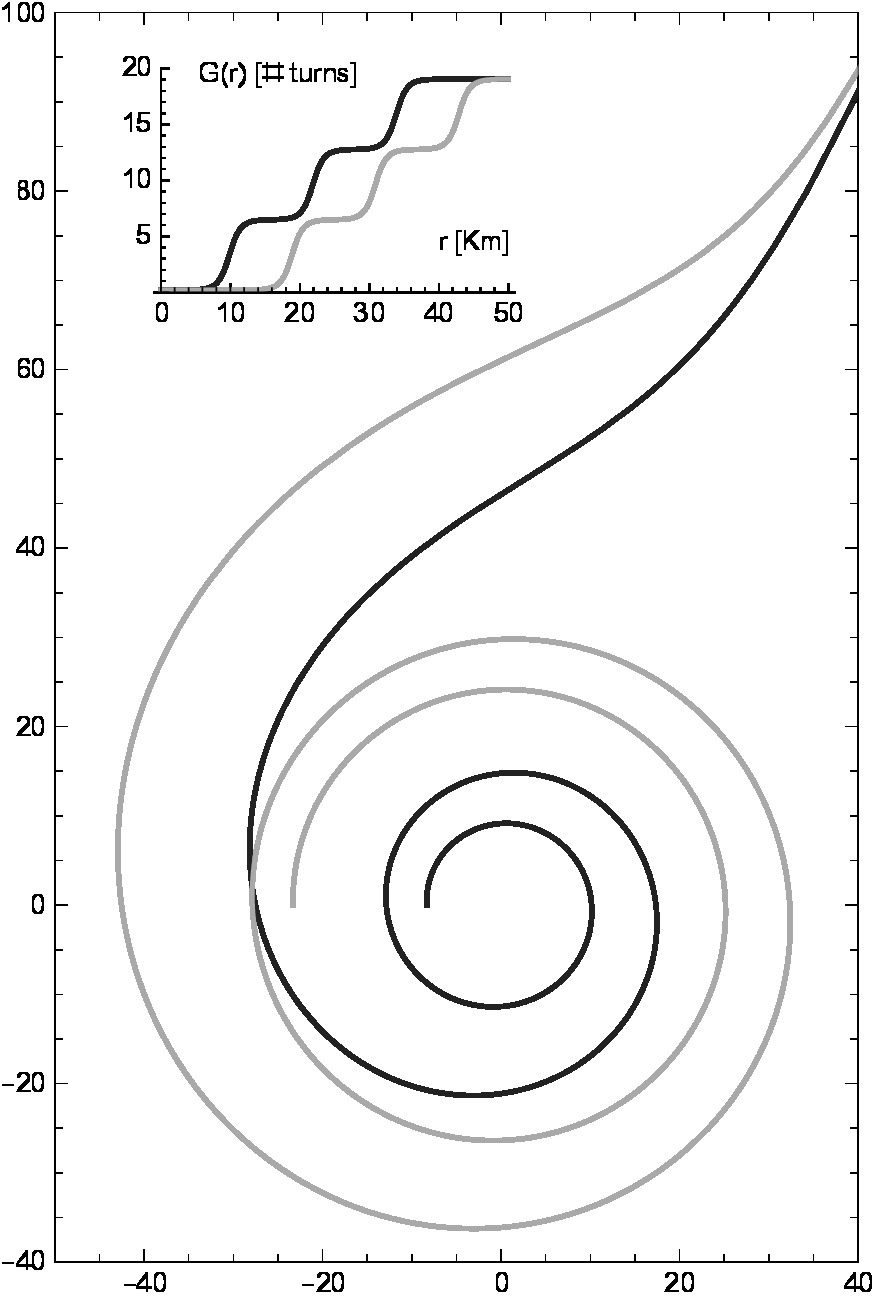} 
	\caption{Example of time evolution of a spiral whose phase $\chi(r,\theta,t)$ is a multiple front kink soliton solution of the sine-Gordon Eq. (\ref{eq.sinegrod}). The scale of the frame is Km. From initially black spiral to  gray spiral the time span is five hours. The corresponding $\theta=G(r,t)$ solution is presented in the upper inset, at the same moments of time.}
	\label{figsolitons}
\end{figure}
with real functions $f, \chi$ such that $f$ varies slowly with the radial distance $r$, while $\chi$ varies quickly. Independently of the procedure to find the exact expression for the right hand term in Eq. (\ref{eq.sine.gord})
we can always expand this term in a Taylor series in all its arguments, around zero. Since the zero and first order terms were already considered in the left hand part, the right hand term begins with quadratic terms, like for example $C_0 \Phi^2$. When we implement the form of solutions from Eq. (\ref{eq.substit}), after some algebra, the imaginary part of the resulting system of equations reads 
\begin{equation}\label{eq.sinegrod}
\chi_{tt} -\triangle_{cyl}\chi =C_0 f \sin (\chi) +\mathit{O}(3)=0,
\end{equation}
which, in the approximation of slow variation of the amplitude $f$ in comparison to the phase $\chi$,  is the sine-Gordon equation \cite{lambj,ablo81,abloBook2}. The remaining real part of the equation is used to determined the function $f$, and the integration of the momentum conservation system of equations determines the velocities. This result is independent of any procedure used to decompose the system of equations, hence it has a very high degree of generality and validity.

Multiple turns spirals are obtained by using multiple-fronts kinks (topological solitons) solutions of the sine-Gordon equation, built by gluing together shifted kinks on plateau of arbitrary width of other base kinks, and so on. Such solutions were obtained in the case of nonlinear dispersion equations, \cite{alt} (sometimes referred to as kovatons, \cite{rose}) whose stability is the same as for regular kinks, since multiple-fronts kinks have the same minimum value of energy as kinks, being calculated from their Hamiltonian (see e.g. \cite{lambj}, pp. 150).

%%%%%%%%%%%%%%%%%%%%%%%%%%%%%%%%%%%%%%%%%%%%%%%%%%%%%%%%%%%%%%%%
%%%%%%%%%%%%%%%%%%%%%%%%%%%%%%%%%%%%%%%%%%%%%%%%%%%%%%%%%%%%%%%%
%%%%%%%%%%%%%%%%%%%%%%%%%%%%%%%%%%%%%%%%%%%%%%%%%%%%%%%%%%%%%%%%

\subsection{Spiral solutions in the quadratic approximation}

In order to study quantitatively the nonlinear solutions and their stability for our model we retained in Eqs. (\ref{eq9}-\ref{eq11}) only the quadratic terms. The final results in this section amply justify this truncation, because we are able to obtain spirals of various geometries that fit very well the observed patterns in this order of approximation. We search solutions for the IFF in the form $\Phi=f(r) \hbox{Exp} [i (G(r)+m \theta+ \omega_{c} t)]$ where $f, G$  are real functions, and we use  complex frequency $\omega_{c}=\omega_r+i \omega_i$ to take into account dissipation too, $m$ is integer, and we denoted like before $c^2 =(P_w / \rho_w -g T_0)$. 
By integrating Eqs. (\ref{eq10}-\ref{eq11}) with respect to time
$$
(1-a \Phi) u=\frac{-i c  (f'+i G'f)}{\omega_c } e^{i(G+m \theta+i \omega_c t)}+u_{s}(r, \theta),
$$
$$
(1-a \Phi) v=\frac{m c f}{\omega_c r } e^{i(G+m \theta+i \omega_c t))}+v_{s}(r, \theta),
$$
and by plugging these velocities in Eq. (\ref{eq9}) we can separate three main terms. The real part of the time dependent terms has the form
\begin{equation}\label{equ.231513a}
G''f+2 f' G'+\frac{G'f}{r}+\frac{2 a \omega_r \omega_i f}{c^2}=0.
\end{equation}
The imaginary part of the time dependent terms has the form
\begin{equation}\label{equ.231513}
f''+\frac{f'}{r}+\biggl[ -G^{'2}-\frac{m^2}{r^2}+\frac{a (\omega_{r}^{2}-\omega_{i}^{2})}{c^2} \biggr] f =0,
\end{equation}
and the time-independent terms are 
\begin{equation}\label{equ.231513b}
(r u_{s})_r+v_{s,\theta}=0,
\end{equation}
where we mention that the $\theta$ derivative here does not reduce to $m$ because the stationary solutions do not generate spiral patterns.
Eqs. (\ref{equ.231513}-\ref{equ.231513b}) form a nonlinear differential system for the  functions $f(r), G(r)$, $u_s (r,\theta)$ and $v_s (r,\theta)$ and from its solutions we can build the fields $\Phi, u, v$ and eventually predict occurrence of sea-ice patterns formed on the water surface from our nonlinear model in the quadratic approximation, Eqs. (\ref{eq9}, \ref{eq10}, \ref{eq11}), or in its the compact version Eqs. (\ref{eq.finals41}, \ref{eq.finals4}, \ref{eq.finals42}).

Eqs. (\ref{equ.231513}-\ref{equ.231513b}) are similar to very well studied NLS-type equations in literature. A very simplified version of our radial model Eq. (\ref{equ.231513}) was derived by Benney and Roskes, \cite{abloBook2,benney69}, for the case of a $(2+1)-$dimensional NLS-type equation for slowly varying envelope surface water waves in potential flow and finite depth. The BR equation can explain long-time similarity solutions of NLS with radiative tails. A more general form for NLS-type integrable system  was obtained by including surface tension effects to the BR equation shallow water, \cite{ddjord,benney69}.   
In addition to general nonlinear integrable models for shallow water waves   using the BR equation, \cite{ablo81}, similar equations to Eq. (\ref{equ.231513}) are also described in nonlinear optics, where they are referred as the Davey-Stewartson (DS) equation, \cite{papa94,ablo2005},  or for $\chi^{(2)}$-nonlinear optics \cite{ablo2001,craso2003}. Moreover, it was shown, \cite{abloBook2,ablo2001}, that the DS equation applied to shallow water waves with large coefficient of surface tension admits localized boundary induced pulse solutions, or weakly decaying lump-type solutions, including the so-called dromions, exhibiting interesting nonlinear behaviors including wave collapse, blow-up  and dark-lump solitons. These results can be somehow condensely expressed through a multidimensional ($d \ge 1$) generalized NLS equation analysis, \cite{sulsul99},  of the form
\begin{equation}\label{eq.nls.gen}
i \Psi_t+\triangle_{d} \Psi+|\Psi|^{2 \sigma} \Psi=0.
\end{equation}
It was shown that Eq. (\ref{eq.nls.gen}) admits subcritical  global solutions  $ \Psi(r,\theta,t)=T(r) \hbox{Exp} [i (G(r)+\omega t)] $, for which blow-up does not occur, if $d \sigma <2$. With this substitution, the resulting equation becomes
\begin{equation}\label{eq.nls.gen2}
T''+\frac{1}{r} T'+\biggl( -G^{'2} -\frac{m^2}{r^2} -\omega\biggr) T+T^{2 \sigma+1}=0,
\end{equation}
which has the so-called Townes modes as pulses solutions, if $2 \sigma \le 2$,  \cite{weinst}. The Townes modes can become stable solitons, \cite{townes},  in the case of self-focusing NLS solution collapse and if their energy is smaller than the critical energy $2 \pi \int_{0}^{\infty} r T^2(r) dr$, \cite{weinst}. In our case, Eq. (\ref{equ.231513}), by neglecting dissipation,  $\omega_i << \omega_r$, the coefficient $2 \sigma \sim 1$ which fulfills the subcriticality condition of existence of global stable solitons.

Eq. \ref{equ.231513b} represents a divergence free condition for the stationary part of the velocity flow, namely $\nabla \cdot (u_s,v_s)=0$, coming from the incompressible mass conservation of the stationary part.   Simplified solutions $f(r$) where $G(r)=\pm m$ would not generate swirl patterns, but rather oscillating monomials in $r$, because such a choice means neglecting a strong nonlinear coupling between the IFF function and the velocities. The term $1-a \Phi (r,\theta,t)$ occurring systematically next to the components of the velocities is not too relevant for the dynamics because the small values   ($a\simeq 0.3$, $\Phi \simeq 0.5$) in front of the time oscillating part under the complex logarithm. The time variation of this term is in second order of smallness, so occasionally the term can be considered constant for rough qualitative evaluations. The consideration of complex frequency helps in the evaluation of the dispersion relations, and in the study of the stability of the nonlinear solutions as it will be seen in continuation. Actually, for zero damping effects $\omega_i=0$ Eq. (\ref{equ.231513b}) becomes identical to the DS (or BR) equations. In this ideal case we expect that the spiral solutions, in appropriately chosen system of coordinates become a representation of Townes modes lump-solitons, or rather lump-breathers \cite{weinst,townes}.

Because the amplitude $f$ and the phase $G(r)$ are intertwined in the system Eqs. (\ref{equ.231513a}, \ref{equ.231513}) we obtain a more convenient version where the amplitude $f$ is decoupled from the phase, and fulfills the equation
$$
f^{''}+\frac{f'}{r}+\biggl( \frac{a (\omega_{r}^{2} -\omega_{i}^{2})}{c^2}-\frac{m^2}{r^2} \biggr)f-\frac{C_{1}^{2}}{r^2 f^3}
$$
\begin{equation}\label{eq.2342343}
+\frac{4 a \omega_r \omega_i C_1}{c^2 r^2 f^3}\int_{0}^{r}z f^2(z) dz=\frac{4 a^2 \omega_{r}^{2} \omega_{i}^{2}}{c^4 r^2 f^3}\biggl( \int_{0}^{r}z f^2(z) dz \biggr)^2.
\end{equation}
where $C_1$ is a constant of integration. Once Eq. (\ref{eq.2342343}) is solved for $f$, we can implement that solution into Eq. (\ref{eq.2342343}) and obtain the phase solution, \textit{i.e.} 
\begin{equation}\label{eq.2356aa}
G(r)=\int_{0}^{r} \frac{C_1 -\frac{2 a \omega_r \omega_i }{c^2}\int_{0}^{w} z f^2(z)dz}{w f^2 (w)}dw,
\end{equation}
such that the coupling is now reduced to the shared constant $C_1$.
Eq. (\ref{eq.2342343}) is a strongly nonlinear integro-differential equation with singularities at $r,f=0$ very difficult to solve and even to characterize qualitatively. Nevertheless, we note that in the initial version of the equation for the amplitude $f$, Eq. (\ref{equ.231513}), the phase enters only as $G'$, so from the phase solution in Eq. (\ref{eq.2356aa}) we notice that the control term towards $r \rightarrow \infty$ is the denominator $r f^2$. If at large distance from its center the spiral pattern is surrounded by ice $(f=1)$ or by a uniform mixture of water and ice, so asymptotically $f  \neq 0$, it results that the phase $G$ approaches a constant asymptotic value, which would generate logarithmic type of spirals. However, if the spiral pattern is surrounded by water and $f \rightarrow 0$ towards the boundaries, the asymptotic behavior of the phase $G$ can be unstable and depends on how fast the amplitude $f$ approaches zero value.
In order to insure spiral stability we need $r f^2 \rightarrow 0$ which requests $f(r \rightarrow \infty) \sim \mathcal{O}(r^{-\alpha})$ with $\alpha \ge 1/2$. Even in this case, the fourth term in Eq. (\ref{eq.2342343}) can approach infinity because of its higher power $f^3$ in denominator. 

In order to avoid this ambiguity in our analysis, we consider here only solutions with $C_1 =0$. In this case Eq. (\ref{eq.2342343}) reduces to its first three terms in the LHS, which are nothing but a Bessel type of Sturm-Liouville linear differential operator, and in the RHS the square of the integral. With the substitutions $r^2 = 2q$ and 
$$
\varphi (q)=\int_{0}^{q} f^2(\sqrt{2q})dq,
$$
we can map the integro-differential equation into a quadratic nonlinear differential equation
\begin{equation}\label{eq.iur2h2il}
\varphi^{'''} \varphi^{'}-\frac{(\varphi^{''})^2}{2}+\frac{\varphi^{''}\varphi^{'}}{q}+\frac{a(\omega_{r}^{2} -\omega_{i}^{2}) (\varphi^{'})^2}{c q}-\frac{m^2 (\varphi^{'})^2}{2 q^2}-\frac{4 a^2 \omega_{r}^{2} \omega_{i}^{2}  \varphi^{2}}{2 c^4 q^2}=0.
\end{equation}
We thus obtained a nonlinear ODE model through Eqs. (\ref{equ.231513a}, \ref{equ.231513}), or equivalently  Eqs. (\ref{eq.2342343}, \ref{eq.2356aa}), or finally more particular system of order 3 through  Eqs. (\ref{eq.2342343}, \ref{eq.iur2h2il}), which generate an exact solution in the quadratic approximation for the IFF field as $\Phi=f \hbox{exp} [i(G(r)+m \theta+(\omega_r +i \omega_{i} )t)]$.
 
In the following we present an example of stable ($\omega_i =0$, no dissipation) Archimedean spiral pattern $G_0 r+m \theta + \omega_r t=$const., as a  asymptotic solution of Eqs. (\ref{equ.231513}, \ref{equ.231513a}). Indeed, for $G(r)= G_0 r$ we have $G^{'2}=G_{0}^{2}$ constant and Eq. (\ref{equ.231513}) provides the solution $\Phi = f \hbox{exp} [i(G_0 r+m \theta + \omega_r t)]$ with
\begin{equation}\label{eq.bess.funct.arch.spir}
f(r)=a_1 J_{m}\biggl(\frac{r}{c}\sqrt{a\omega_{r}^{2}-G_{0}^{2}c^2}\biggr)+a_2 Y_{m}\biggl(\frac{r}{c}\sqrt{a\omega_{r}^{2}-G_{0}^{2}c^2}\biggr),
\end{equation}
where $a_{1,2}$ are constants. In the asymptotic region, for large enough $r$, regime which is very easy to meet given the space extension of the spiral solutions, the Bessel functions in the solution Eq. (\ref{eq.bess.funct.arch.spir}) behave like $f\rightarrow 1/\sqrt{r}$, which makes $G' \sim 1/(r f^2)$ a constant, exactly as it should be to close the system. In a similar way we find a solution generating a logarithmic spiral $G1 \log r+ m \theta + \omega_r t=$const. By implementing $G^{'2}=G_{1}^{2}/r^2$ in Eq. (\ref{equ.231513}) generates the solution $\Phi = f \hbox{exp} [i(G_1 \log r+m \theta + \omega_r t)]$ with amplitude
\begin{equation}\label{eq.bess.funct.log.spir}
f(r)=a_0 + a_1 J_{m}\biggl(\frac{r \omega_r}{c} \sqrt{a(G_{1}^{2}+m^2)}\biggr)+a_2 Y_{m}\biggl(\frac{r \omega_r}{c} \sqrt{a(G_{1}^{2}+m^2)}\biggr).
\end{equation}
In this later case, in the asymptotic region for large $r$ the amplitude approaches the constant $a_0$ and the expression $1/(r f^2)$ approaches $G' \sim G_1 /r$ as requested.
\begin{figure}
	\centering
	\begin{tabular}{cc}
	\includegraphics[width=6cm,height=6.6cm]{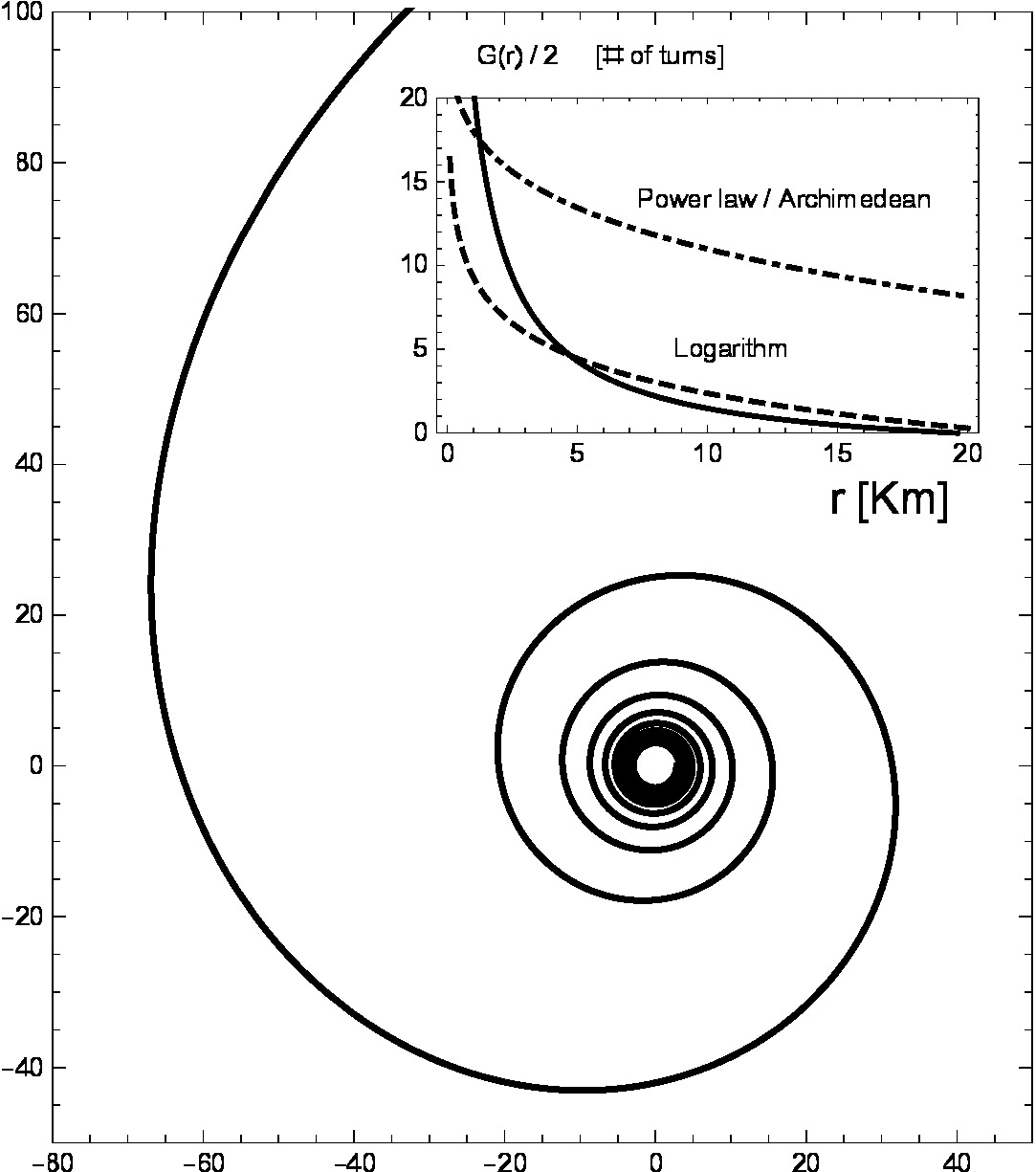} &
	\includegraphics[width=4cm,height=6.6cm]{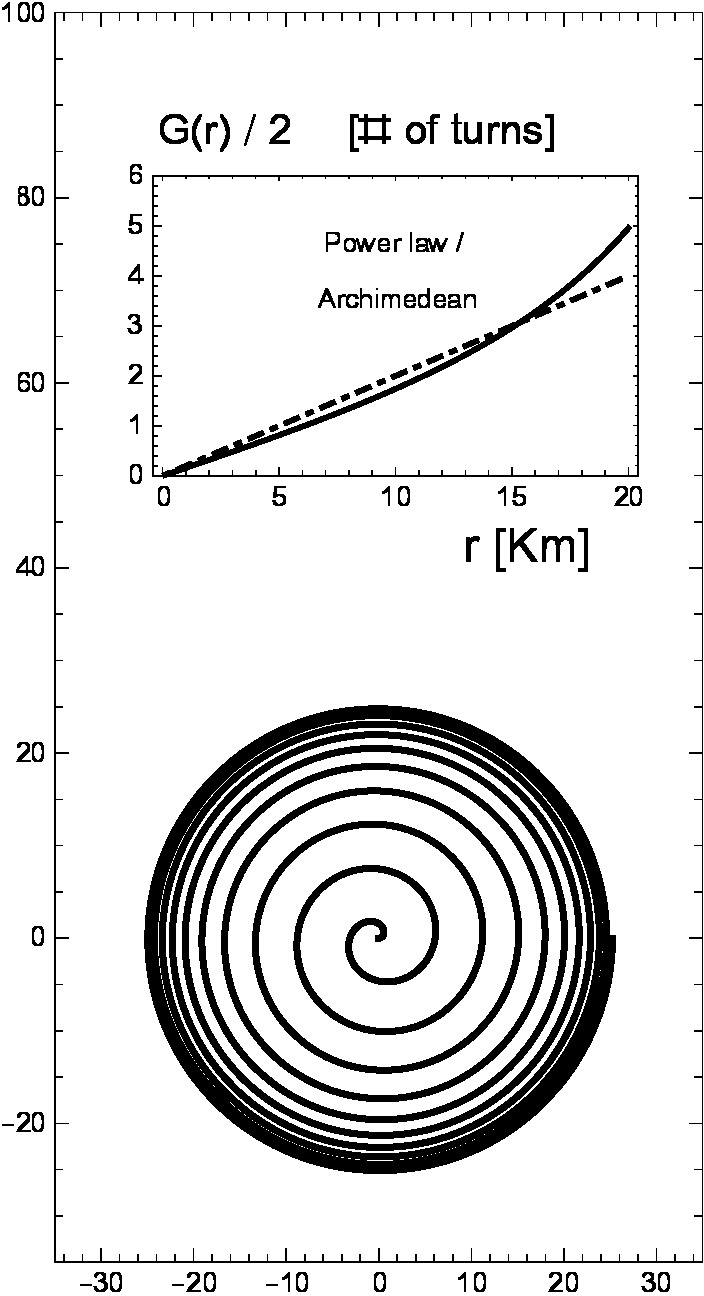} 
\end{tabular}	
	\caption{Spiral shapes as locus of points in plane with $\Phi=1$ (ice crests) obtained with numerical integration of Eqs. (\ref{eq.2356aa}, \ref{eq.2342343}). The resulting equation of the spiral itself is plotted in the inset, together with the generic equations for Archimedean and logarithmic spirals, for comparison.}
	\label{figspiraspira}
\end{figure}
We close this section by presenting an exact IFF amplitude $f$ (or $\varphi$) solution for the third order nonlinear Eq. (\ref{eq.iur2h2il}) in the non-dissipation case, $\omega_i =0$. The absence of dissipation involves the dropping of the integral term in this equation, or equivalently dropping of the non-derivative term in the $\varphi$ equation.
$$
f(r)= \hbox{Exp} \biggl\{ \int_{0}^{r} \frac{a_1 \Gamma(1-m)\biggl[ J_{-m-1}\biggl( \frac{r\omega_r \sqrt{a}}{c}   \biggr) -J_{1-m}\biggl( \frac{r\omega_r \sqrt{a}}{c}   \biggr)   \biggr]\frac{\sqrt{a} \omega_{r}}{ 2c}}{ 	 \biggl[ a_1 \Gamma(1-m) J_{-m}\biggl( \frac{r \omega_r \sqrt{a}}{c}  \biggr) + \Gamma(1+m) J_{m}\biggl(\frac{r \omega_r \sqrt{a}}{c}  \biggr)\biggr]} dr \biggr\}
$$
\begin{equation}\label{eq.sol.varphi}
\times \hbox{Exp} \biggl\{ \int_{0}^{r} \frac{\Gamma(1+m)\biggl[ J_{m+1}\biggl(\frac{r\omega_r \sqrt{a}}{c}  \biggr) -J_{m-1}\biggl( \frac{r\omega_r \sqrt{a}}{c} \biggr) \biggr]\frac{\sqrt{a} \omega_{r}}{2 c}}{ \biggl[ a_1 \Gamma(1-m) J_{-m}\biggl( \frac{r\omega_r \sqrt{a}}{c}  \biggr) + \Gamma(1+m) J_{m}\biggl( \frac{r\omega_r \sqrt{a}}{c}  \biggr)\biggr]} dr +a_0 \biggr\},
\end{equation}
$a_{0,1}$ being constants and with the phase obtained by implementing this amplitude $f$ in Eq. (\ref{eq.2356aa}). One example of such solution is presented in Fig. \ref{figspira1}.
\begin{figure}
	\centering
	\includegraphics[width=7cm,height=7cm]{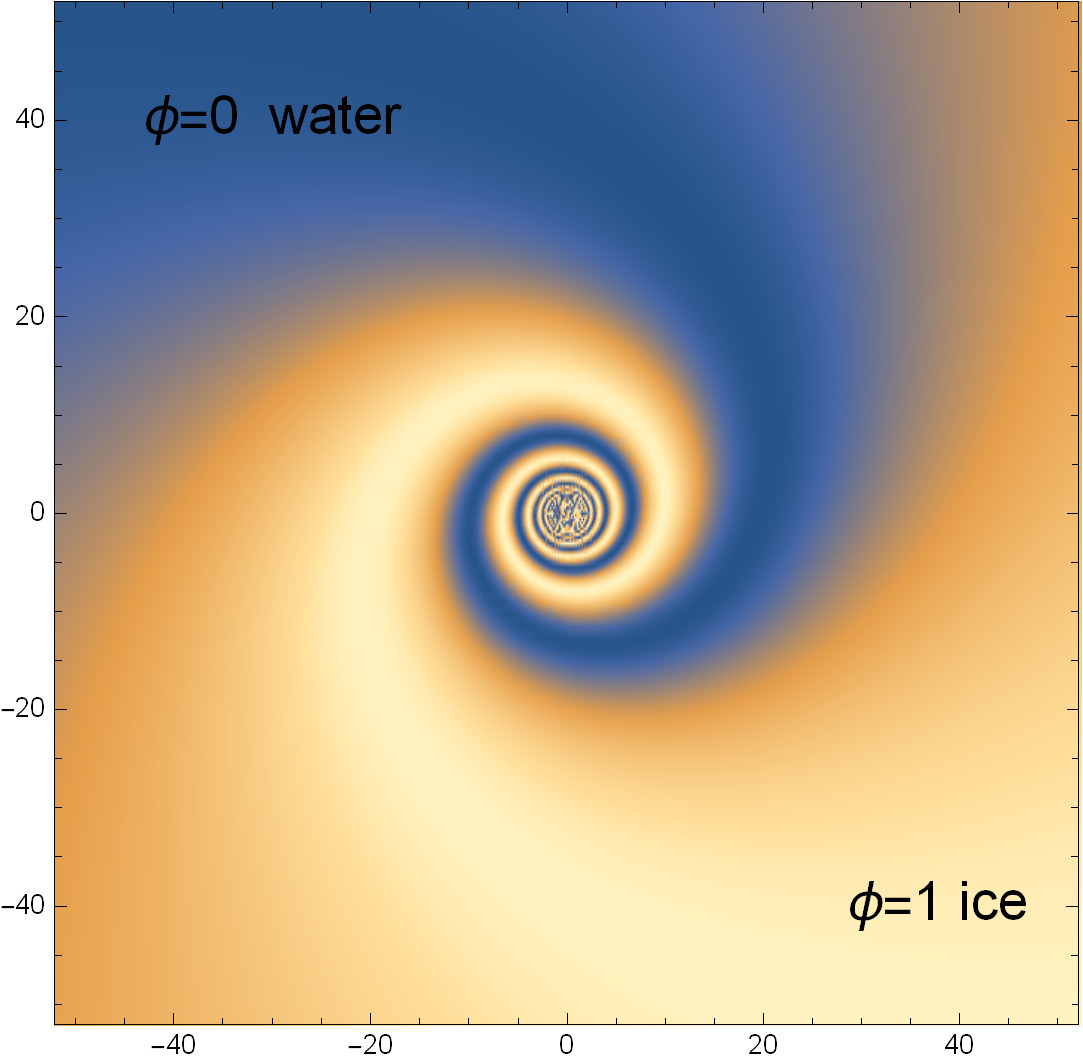} 
	\caption{Density plot of the IFF function $\Phi(r,\theta)$ as a logarithmic spiral solution obtained from Eq. (\ref{eq.sol.varphi}).}
	\label{figspira1}
\end{figure}
In addition to the specific spiral geometries, the present model can predict the average size of the sea-ice spiral patterns, as well as the number of turns, or their pitch. The general equation of a time independent spiral can be written in the general form $\cos [G(r)+m \theta]$ or $\cos [r+m G^{-1}(\theta)]$ where cosine or any other periodic function can be used, the integer $m$ is the number of arms, its sign identifies the helicity of the spiral. The spiral equation in polar coordinates becomes $r=\pm m G^{-1}(\theta)$. For one arm spirals, the pitch is given by $p=G^{-1}(\theta+2 \pi)-G^{-1}(\theta)$. For example, the Archimedean spiral $\alpha r+ \theta $ has $p=2 \pi /\alpha$, and the logarithmic spiral $\alpha \ln r+ \theta$ has variable pitch $p=r(e^{2 \pi / \alpha}-1)$. Independent of the mathematical approach of the model equations, fully nonlinear analysis, quadratic approximation, linearized, etc., we notice that the solutions for the IFF distributions are always combinations of Bessel functions, see for example Eqs. (\ref{eq21},\ref{eq32345},\ref{eq.psipsi2},\ref{eq.bess.funct.arch.spir},\ref{eq.bess.funct.log.spir},\ref{eq.sol.varphi}), consequence of the basis generated by the linearized equations. All solutions contain the general term 
$$
\Phi \sim A J_{m}(\alpha r , \dots) \rightarrow_{r\rightarrow \infty } \frac{A}{\sqrt{2 \pi \alpha r}} \cos (\alpha r+\cdots),
$$
where $\alpha$ stands generically for the coefficient in front of $r$ for various solutions. 
From Eq. (\ref{eq.2356aa}), and considering just  the dominant term in the asymptotic region (where the large part of the spiral actually unfolds) for the phase $G(r)$ function, we can write 
$$
G(r) \sim \int^{r} \frac{C_1}{r |\Phi(r,\dots)|^2}dr  \rightarrow_{r\rightarrow \infty } \frac{2 \pi C_1}{A^2} \tan (\alpha r),
$$
which implies the following spiral parameterization 
$$
r=G^{-1}(\theta)\sim \frac{1}{\alpha}\arctan \biggl( \frac{A^2 \theta}{2 \pi C_1} \biggr).
$$
Since $A$ is of the order of unity because of the range for $0 \le \Phi \le 1$, and stable solutions have usually $C_1$ negligible and very small, the limiting radius of  such spirals is given by $r_{max}=L=\pi / (2 \alpha)$, see Fig. \ref{figspiraspira} right, and the pitch is almost constant within the whole active range of $\theta$,  $p \simeq \pi/\alpha$. From the linearized model, Eq. (\ref{eq24}), we know the scaling factor in front of $r$ inside the Bessel functions is $\xi_{m,n}/L$. If we consider the fundamental mode $m=1$, which is probably the most stable, knowing that $\xi_{1,n} \sim 3.83 n$ we can consider the mean value of the pitch $p=L/\mathcal{N}$, where $\mathcal{N}$ is the mean number of full turns of such a spiral. Hence we have $\alpha \sim \xi_{1,n} /L \sim \pi / p \sim \pi \mathcal{N} / L$ where from $\mathcal{N} \sim \xi_{1,n}/ \pi \sim 1.2 n$. It results the spiral have at least one full turn, and the maximum number of turns is very close the the highest order term taken in the Bessel-Fourier series for the IFF solution. Moreover, for ice fragments with average draft $T_0 \simeq 10$m, and water pressure upon ice blocks in the range of normal atmospheric pressure, we can evaluate from Eqs. (\ref{eq19}, \ref{eq25}) and from $\alpha \sim \sqrt{a} \omega_{r} /c $ that $0.25 \le 2 \pi \alpha/\omega_r =\alpha T  \le 0.45$. This relation together with $L \sim \xi_{1,n}/\alpha $ gives a fundamental general criterion to predict the size (and from here the velocity, energy and vorticity) of such spirals in relation to their period $T$ of revolution. For example a spiral with a period of revolution ten hours can be modeled by our solution with the scaling factor  in from of $r$ variable  $\alpha \sim 5 \times 10^{-4}$m$^{-1}$ predicting a spiral/swirl size of $L \sim 30 \div 40$Km. Moreover, the parameter $\alpha T$ shown above provides a sort of inverse of characteristic velocity of the spiral.

In order to connect the spiral evolution with possible solitary waves we start again from the quadratic approximation Eqs. (\ref{equ.231513a}, \ref{equ.231513}) and,  in the case of weak dissipation such that we can  neglect  lower order terms containing the product $\omega_r \omega_i$, and by using the substitutions 
$$
\xi=\ln r, \ \ S(\xi)=f^2(e^{\xi}), \  \ \ H(\xi)=\int^{e^{2 \xi}}f(\sqrt{z}) dz.
$$
we can transform  Eq. (\ref{equ.231513}) in a nonlinear, variable coefficients integro-differential equation
$$
\frac{S^{'2}}{8}-\frac{m^2 S^2}{2}+b_1 S+\frac{a (\omega_{r}^{2}-\omega_{i}^{2})}{2 c^2} e^{2 \xi}S^2-\frac{a (\omega_{r}^{2}-\omega_{i}^{2})}{c^2}S \int^{\xi}S e^{2z}dz
$$
\begin{equation}\label{eq.7yhthth}
+\frac{2 a b_2 \omega_r \omega_i}{c^2}S\int^{\xi}\frac{H S'}{2 S^2}dz-\frac{a^2 \omega_{r}^{2} \omega_{i}^{2}}{2 c^2}S\int^{\xi}\frac{H^2 S'}{S^2}dz=b_3,
\end{equation}
where $b_{1,\dots,3}$ are constants of integration. By further substituting $T(\xi)=\ln S(\xi)$ we can further  reduce  Eq. (\ref{eq.7yhthth}) to a partially separated form
\begin{equation}\label{eq.37372634}
T^{'2}+\mathcal{F}_1 (T)+\mathcal{F}_2 (\xi)+\mathcal{I}[T]=0,
\end{equation}
where $\mathcal{F}_{1,2}$ are functions, and $\mathcal{I}$ is nonlinear integral operator. The first three non-integral terms in Eq. (\ref{eq.37372634}) form a known differential form which accepts exact solutions in parametric form (see for example \cite{odebook} equation 13.6.3.8). Consequently, this modified form obtained for the model equation can be approached by integral equations technique to analyze existence and uniqueness of solutions. 

The advantage of such a change of variables through the logarithmic substitution may favor the representation of solutions in terms of traveling localized perturbations along the $\xi$ axis.

%%%%%%%%%%%%%%%%%%%%%%%%%%%%%%%%%%%%%%%%%%%%%%%%%%%%%%%%%%%%%%%%%%%%%%%%%%%%%%
%%%%%%%%%%%%%%%%%%%%%%%%%%%%%%%%%%%%%%%%%%%%%%%%%%%%%%%%%%%%%%%%%%%%%%%%%%%%%%
%%%%%%%%%%%%%%%%%%%%%%%%%%%%%%%%%%%%%%%%%%%%%%%%%%%%%%%%%%%%%%%%%%%%%%%%%%%%%%
%%%%%%%%%%%%%%%%%%%%%%%%%%%%%%%%%%%%%%%%%%%%%%%%%%%%%%%%%%%%%%%%%%%%%%%%%%%%%%

\section{Nonlinear Stability}
\label{secstab}

The surface of the sea is a random nonlinear wave field, and coherent structures localized in time and space such as wave trains, rogue waves, and wave packages also get birth randomly. There is no correlation between waves at large distances except in cases where wind would blow constantly and uniformly on a large area, but even these events are transient and localized in time. For example, even in the case of the largest wave phase velocity (20 m/s) it would take about half an hour for information to travel from an initial point at the water surface, along 30 Km, time during which the conditions at the initial place have changed already, since the transient seas can change in matter of minutes. Studies on correlations of surface waves back up this conclusion  \cite{stabi1}.
The case becomes even more complicated when there are other systems in the water which interact with waves and currents, like ice fragments, oil spots, garbage, algae, etc. It is exceptional when a large space and time scale stable structure occurs spontaneously at the surface of the sea, like the spirals observed in the polar seas, \cite{1,2}, Fig. \ref{fignews}. Such spiral structures extend over tens of kilometers, move very slow and survive for days. Such stable, large space-scale, long life time rotating patterns must be generated by a strong type of correlation and nonlinear coupling of modes. The ice and water spirals observed, \cite{1,2}, are definitely the signature of coherent, nonlinear collective interaction, generated by  spontaneous large-scale nonlinear collective sea modes.
 
The linear solutions $\Phi^{1}_{m,n}$ given by Eqs. (\ref{eq25}-\ref{eq27}) form a complete orthonormal basis of functions for the continuous functions defined on $[r_0,L] \times [0, 2 \pi]$, which means that any solution of the fully nonlinear system Eqs. (\ref{eq14}-\ref{eq16}), or any solutions of some lower order of  approximation containing quadratic, cubic, etc. nonlinear terms can be expanded in this basis.

In this section we consider nonlinear terms up to the second order of approximation, which of course includes the linearization Eqs. (\ref{eq17}-\ref{eq18}), already analyzed in section \ref{sec.lin.sol}. The system Eqs. (\ref{eq9}-\ref{eq11}), written in dimensionless form in the second order of approximation with the terms slightly re-ordered, contains now cubic nonlinearities, and  has the form    
\begin{equation}\label{eq35}
(1-a \Phi) \biggl( u_{\zeta} +\frac{u}{\zeta} +\frac{v_{\theta}}{\zeta} \biggr) -a ( a_1 \Phi_{\tau} +\delta u \Phi_{\zeta} ) =0, 
\end{equation}
\begin{equation}\label{eq36}
(1-a \Phi) ( a_1 u_{\tau} + \delta u u_{\zeta} ) -a ( a_1 u \Phi_{\tau} +\delta u^2 \Phi_{\zeta} )+a_2 \Phi_{\zeta}= 0,
\end{equation}
\begin{equation}\label{eq37}
(1-a \Phi) \biggl( a_1 v_{\tau} + \delta u v_{\zeta} +\delta \frac{uv}{\zeta} \biggr) -a ( a_1 v \Phi_{\tau} +\delta u v \Phi_{\zeta} ) +\frac{a_2 \Phi_{\theta}}{r}= 0,
\end{equation}
where we denoted 
$$
a_1 =\frac{\varepsilon R }{\delta \Theta}, \ \ \  a_2 =\frac{\varepsilon \Theta \biggl( \frac{P_w}{\rho_w}- g T_0 \biggr) }{\delta R} .
$$
Before applying a rigorous approach to the nonlinear stability of the exact solution for the linearized system, Eqs. (\ref{eq25}-\ref{eq27}), we make some qualitative comments. From the $\theta$ and $t$ dependence of the solution  Eq. (\ref{eq25}), and knowing that the tangent velocity is usually about one order  of magnitude less than the radial velocity, we make the hypothesis $v_{\theta} = \omega v_t$, with $\omega$ a parameter to be adjusted conveniently. The only term where $v$ occurs in Eqs. (\ref{eq35}-\ref{eq37}) is through $v_{\theta}/r \sim \omega v_{t}/r$. If we perform this substitution in Eq. (\ref{eq35}) the resulting term has a smaller order compared to the rest of the terms in the equation, so it is enough to forward use the linear substitution given by the second equation in Eqs. (\ref{eq18}).   
In this way the sub-system made by the first two equations, Eqs. (\ref{eq25}-\ref{eq26}), depends only on $\Phi$ and $u$, and once a solution is found for these two quantities, we can solve Eq. (\ref{eq37}) for $v$, because this equation transforms now into a linear, first  order PDE in $t$ and $r$ which is fully integrable. Under the above hypothesis Eq. (\ref{eq35}) becomes
\begin{equation}\label{eq38}
(1-a \Phi) \biggl( u_{\zeta} +\frac{u}{\zeta} -a_3 \frac{\Phi_{\theta}}{\zeta} \biggr) -a ( a_1 \Phi_{\tau} +\delta u \Phi_{\zeta} ) =0, 
\end{equation}
with  
$$
a_3=\frac{\varepsilon \Theta \biggl(\frac{P_w}{\rho_w}-g T_0 \biggr) }{\delta R (1-a \Phi^{0})}.
$$
We note that  at large distance from the origin $r\rightarrow \infty$ the terms containing $1/r$ can be neglected in Eq. (\ref{eq38}). After some algebra we obtain for this asymptotic limit the condition
$(1-a \Phi^0 ) u_t =0$ which involves the that the radial speed becomes uniform asymptotically. Consequently, from Eq. (\ref{eq36}), and in the same limit we obtain that all functions $\Phi, u$ and $v$ approach constant and uniform values in the far asymptotic region.

In the following we plug the linear analytic solutions Eqs. (\ref{eq25}-\ref{eq27}) in the nonlinear system Eqs. (\ref{eq36}-\ref{eq38}) containing terms up to order two in $\varepsilon$ and $\delta$. The result is that all linear terms will contain a double series over the order $m$ of the Bessel functions, and over the order $n$ of their zeroes. The quadratic nonlinear terms  given be the second parenthesis left hand side of Eqs. (\ref{eq36}-\ref{eq38}), will involve products of such series. Consequently, the linear solution remain stable with the inclusion of the nonlinear terms if the products of series can be expressed in terms of the solutions series. Namely, if the following relation exists
$$
\sum_{m_1 , m_2=-\infty}^{\infty}\sum_{n_1 , n_2 =1}^{\infty} C_{m_1, n_1 } C_{m_2, n_2} H_{m_1}\biggl( \frac{\xi_{m_1 , n_1 }}{L}r \biggr) H_{m_2}\biggl( \frac{\xi_{m_2 , n_2}}{L}r \biggr) \times
$$
$$
\times  \hbox{Exp}\biggl[i (m_1 +m_2 ) \theta + \frac{ic}{L}t(\xi_{m_1 ,n_1}+\xi_{m_2,n_2})  \biggr]
$$
\begin{equation}\label{eq39}
=\sum_{m=-\infty}^{\infty}\sum_{n=1}^{\infty} C_{m, n} H_{m}\biggl( \frac{\xi_{m, n }}{L}r \biggr) \hbox{Exp}\biggl[i m \theta + \frac{ic}{L}(\xi_{m ,n}t)  \biggr],
\end{equation}
where $C_{m,n}$ are the coefficients of the solution Eq. (\ref{eq25}). Obviously, because of completeness of the Fourier series we have the constraint $m=m_1 + m_2$. This constraint transform the above relation into
$$
\sum_{m_1 =-\infty}^{\infty}\sum_{n_1 , n_2 =1}^{\infty} C_{m_1, n_1 } C_{m-m_1, n_2} H_{m_1}\biggl( \frac{\xi_{m_1 , n_1 }}{L}r \biggr) H_{m-m_1 }\biggl( \frac{\xi_{m- m_1  , n_2}}{L}r \biggr) \times
$$
$$
\times  \hbox{Exp}\biggl[ \frac{ic}{L}t(\xi_{m_1 ,n_1}+\xi_{m-m_1 ,n_2})  \biggr]
$$
\begin{equation}\label{eq40}
=\sum_{m=-\infty}^{\infty}\sum_{n=1}^{\infty} C_{m, n} H_{m}\biggl( \frac{\xi_{m, n }}{L}r \biggr) \hbox{Exp}\biggl[ \frac{ic}{L}(\xi_{m ,n}t)  \biggr].
\end{equation}
The completeness of the Fourier-Bessel series determines the range of the summations over $m_1$ and $m$, and then coefficients $C_{m,n}$ should fulfill the resulting nonlinear relation.  

In the following we apply Arnold's convexity method, \cite{stabi}, to prove stability estimates for smooth solutions in our nonlinear two-phases model. This estimation provides $L^2$ squared integrable
bounds on perturbations of the IFF function and velocities. Our first observation is that the original system Eqs. (\ref{eq6}, \ref{eq7}) , in the inviscid case ($\eta=0$), with density and pressure given by Eqs. (\ref{eq3},\ref{eq4}) is a planar ideal barotropic fluid, and consequently Hamiltonian. Indeed, the pressure 
$$ 
P(\rho)=P_w+\frac{(\rho-\rho_w) \biggl( \frac{P_w}{\rho_{w}}-g T_0 \biggr)}{a},
$$
depends only on the density and thus warrants the barotropicity. From here we calculate the specific enthalpy $h(\rho)$, and specific energy $e(\rho)$ of our fluid by using \cite{thermo}
\begin{equation}\label{eqtherm1}
\nabla h=\frac{\nabla P}{\rho}, \ \ \frac{de}{d\rho}=h,
\end{equation}
and obtain
\begin{equation}\label{eqtherm2}
e=\frac{ \frac{P_w}{\rho_{w}}-g T_0 }{a}(\rho+\rho \ln \rho)+C_1 \rho+C_2,
\end{equation}
with $C_{1,2}$ arbitrary constants of integration. Because the flow is planar and barotropic, and we can express the flow vorticity as $\vec{\Omega}=\vec{k}(v_{2,x}-v_{1,y})$, it results the conservation of the quantity $\Omega/\rho$. Consequently, the volume ($\rho dx dy$) double integral on the plane of any smooth function of real variable $F$ depending only on  this invariant, can act as a Lagrangian multiplier in sum  to the  Hamiltonian of the flow
\begin{equation}\label{stab1}
H=\iint_{\mathbf{R}^2}\biggl[ \frac{\rho v^2}{2}+e(\rho) \biggr] dxdy+\iint_{\mathbf{R}^2} \rho F\biggl( \frac{\Omega}{\rho}\biggr) dxdy,
\end{equation}
given here by the sum between the kinetic and internal energy densities. We consider the linear solutions  of the linearized system obtained in section \ref{sec.lin.sol} for this two-phase, two-dimensional, barotropic flow as equilibrium states. For such equilibrium states the gradient vectors $\nabla [(u)^2 /2+(v)^2 /2 +h(\rho)]$ and $\nabla (\Omega /\rho)$ can be shown to be orthogonal to the velocity \cite{stabi}. Consequently, these gradient vectors must be collinear and hence there must be a functional relation between their potentials
\begin{equation}\label{stab2}
\frac{(u)^2 +(v)^2 }{2}+h(\rho)=G\biggl( \frac{\Omega}{\rho} \biggr) .
\end{equation}
The functional relationship through the function $G$ represents Bernoulli’s law, and $G$ is called the Bernoulli function. This function qualifies for the role of the Lagrangian multiplier density function $F$ from Eq. (\ref{stab1}). By applying to the functional in Eq. (\ref{stab1}) the nonlinear Lyapunov stability criterion, \cite{stabi,dd}, we obtain the following conditions for the stability against small perturbations of our linear solutions 
\begin{equation}\label{stab3}
e^{''}(\rho)-\frac{(u)^2+(v)^2}{\rho}> 0,
\end{equation}
and
\begin{figure}
	\centering
		\includegraphics[width=4cm,height=4cm]{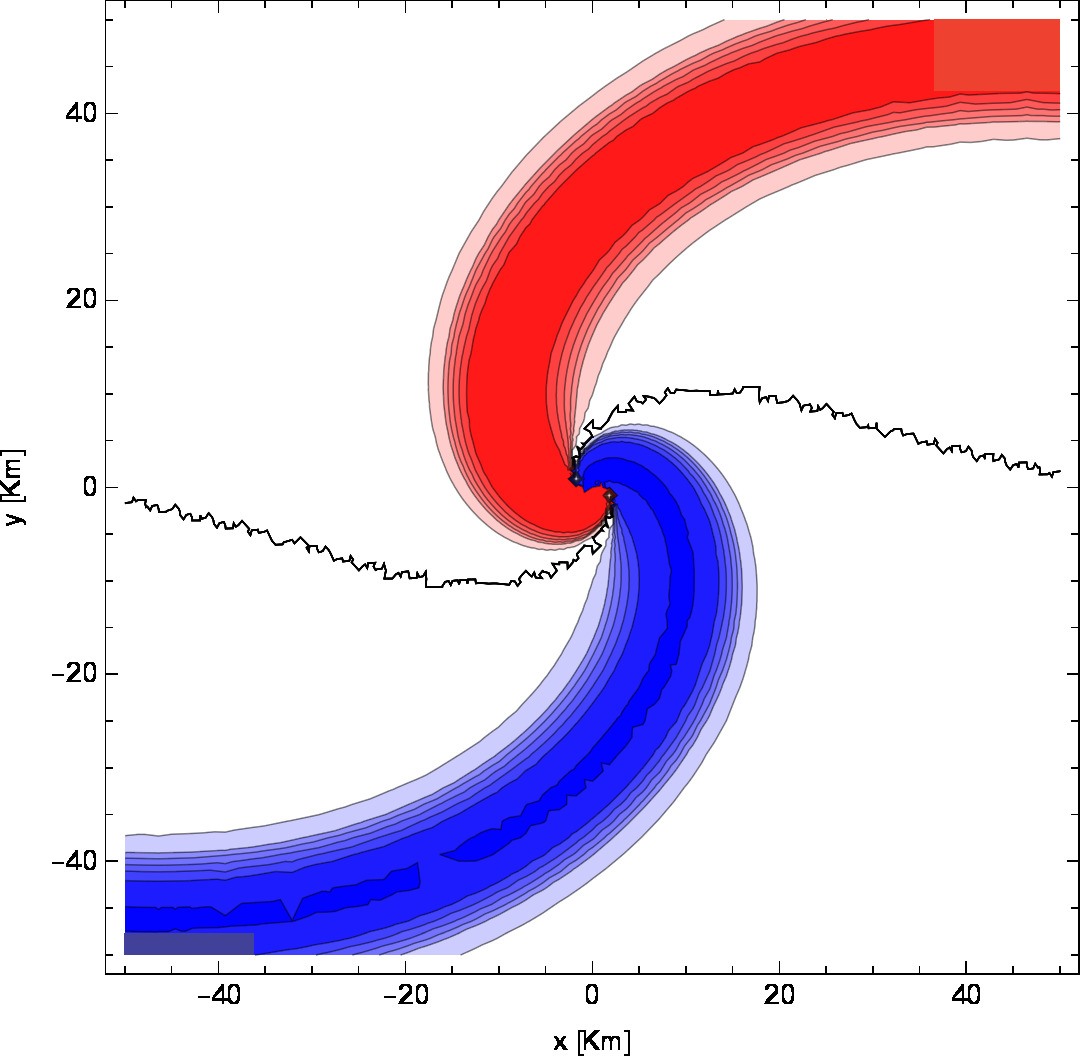} 
	\caption{Fully nonlinear system stability diagrams for large and slow spinning spiral solutions Eq. (\ref{eq.solutnl}) by using Arnold convexity criterion  Eq. (\ref{stab4}). Red represents positive gradient ratio, stable region, bounding the sides of the logarithmic spiral solution, blue represents negative, unstable region, and the black curve stands for zero gradient ratio, i.e. saddle equilibrium separatrix. Inside the blue region of instability ice is not likely to accumulate. The  parameters are $m=1, g_0 =1, f_0 =1, q_0 =60$ and  $r_0 =0.5$ resulting in a fast spinning ($v\sim 50$cm/s) large spiral ($120$Km diameter) with large pitch and wide arms.}
	\label{figarnold12}
\end{figure}
\begin{figure}
		\centering
		\includegraphics[width=3.2cm,height=3.2cm]{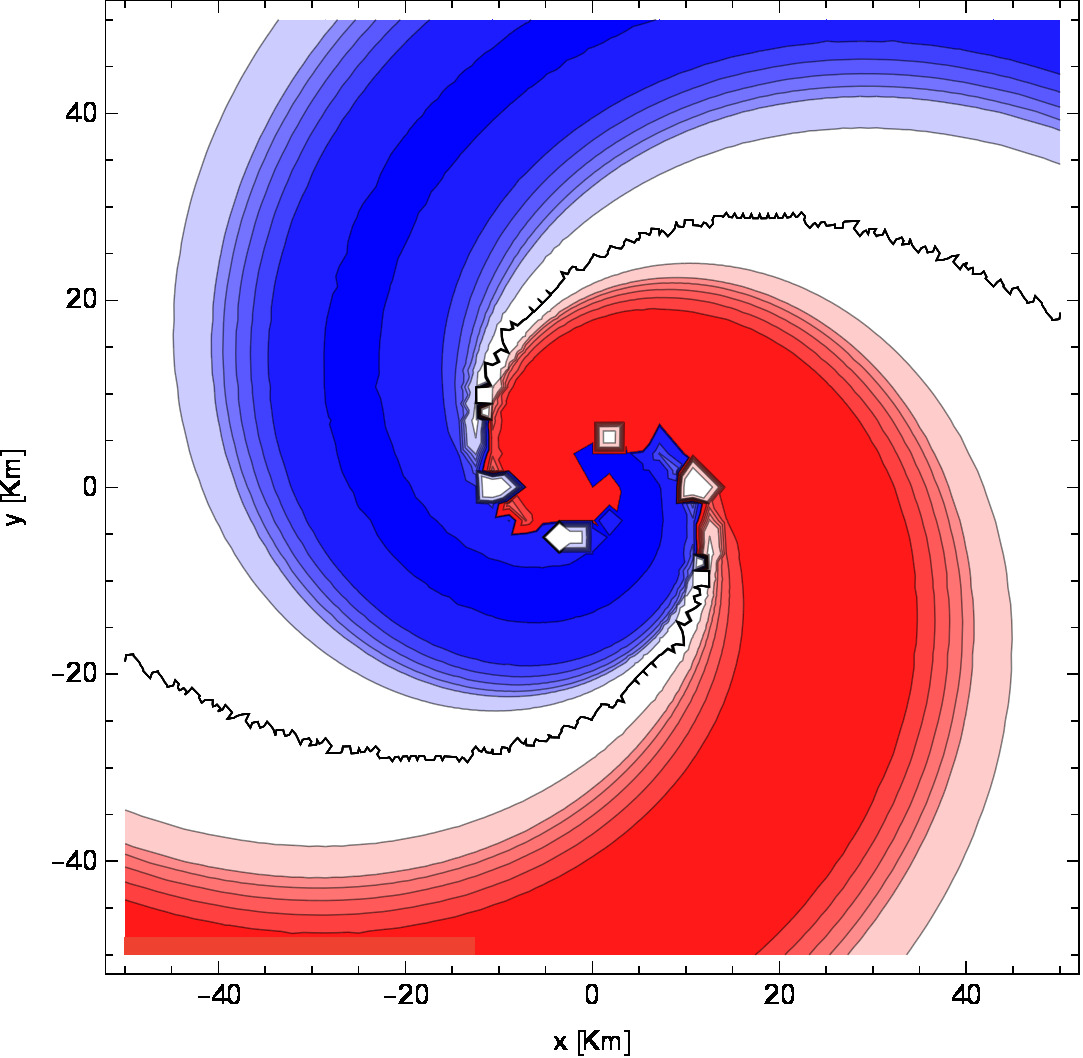} 
	\caption{Same stability diagram as in Fig. \ref{figarnold12} for medium size spirals, diameter $40 \div 80$ Km. Parameters $(r_0 , g_0 , v)= (0.5, 2, 25\hbox{ cm/s})$. The faster rotating the spiral, the narrower arm width has.}
	\label{figarnold345}
\end{figure}
\begin{equation}\label{stab4}
\frac{\nabla \biggl(\frac{(u^1)^2+(v^1)^2}{2}+h(\rho^1) \biggr)}{\nabla \biggl( \frac{\Omega^1}{\rho^1}\biggr)^2}> 0.
\end{equation}
The first stability condition Eq. (\ref{stab3}) is just a request that the flow must be subsonic, because 
it actually reduces to  $(c^2-||\vec{v}^1||^2)/\rho^1 >0$, and this condition is obviously accomplished in the slow motion of water around the sea-ice. The stability constraint arises from the second stability condition,  Eq. (\ref{stab4}). 
\begin{figure}
\centering
\includegraphics[scale=.25]{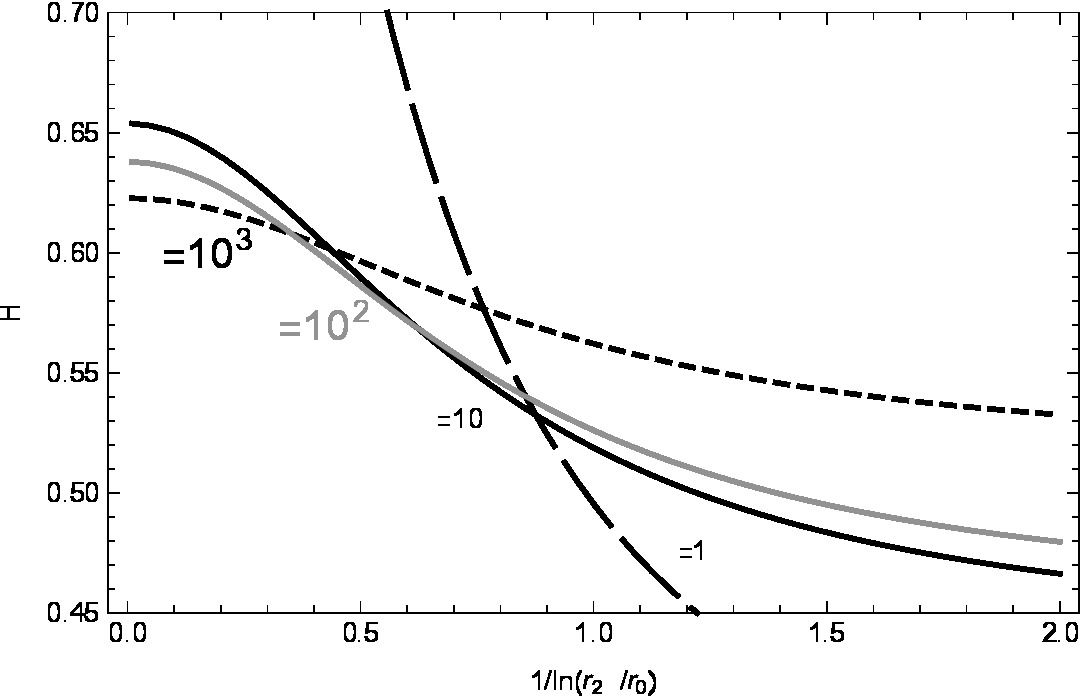} 
\caption{Dependence of the total energy of the spiral system in relative units $H$, Eq. (\ref{stab1}), vs. parameter $g_0$ which is inversely proportional to the  spiral pitch. The four different curves are plotted for various values of the kinetic coefficient $\chi=v^2 g T_0 \rho_w / P_w$, namely the ratio between centrifugal kinetic energy density and water excess pressure upon ice. Since $H$ acts like the free energy of the ice-water, the system follows the minimum energy: if the geometry of the spiral changes (through $g_0$) the system reaches minimum energy through phase transition between different solutions with different $\chi$ values. When a spiral tends to expand, its azimuthal velocity increases, while the water pressure upon ice tends to decrease.}
\label{fighamilt}
\end{figure}
In this equation, it makes sense to write a ratio of gradients, because they are collinear. We calculated the IFF function $\Phi^1$, the density, the corresponding velocity components, their vorticity, and their enthalpy $h$ for $m=1$. In Figs. \ref{figarnold12}, \ref{figarnold345} we plotted the functional from Eq. (\ref{stab4})  for $m=1$ and for values of the parameters that match the geometrical parameters of the swirls observed in the ocean.

\section{Conclusions}
\label{secconcl}

In the present work we investigate a  two-dimensional compressible Navier-Stokes hydrodynamic model design to explain and study large scale ice swirls formation at the surface of the ocean. The regular motion of the ocean water and sea-ice is powered by winds, ocean currents, thermal convection, Coriolis effect. For the mixture ice and water which resides mainly at the surface, the thermal effects produced by internal convection is not a major cause. The observed swirls ot the surface of polar oceans have very large size yet they move very slow which eliminates the Coriolis force as a dominating cause.  Also, the long life time of these structures demonstrates that their formation is rather governed by internal dynamics and ocean currents than waves and  wind. The difference in density between ice and water makes this combination a compressible two-dimensional type of fluid mixture. Such two-dimensional structures with rotational symmetry as noticed by aerial photography  may exercise three types of large amplitude motions: radial oscillations of compression and dilation, azimuthal motion including rotation and shear flow, and a third one consisting in the coupling of radial modes with azimuthal modes and may thus generate very large scale spiral patterns. The question is if indeed these modes tend to couple. Following day and night cycles of warming and cooling the surface ice may melt and freeze back even while the temperature is almost constant (by transfer of heat through radiation, wind and ocean currents). When melting, the ice trapped in the center induces a radial global two-dimensional compression mode. By  the Kelvin theorem, this reduction in radius of the two-dimensional layers, and the corresponding  conservation of circulation induces an increasing in azimuthal motion of the fluid. So, compression induces rotation, which is exactly the requested coupling. The circulation-conserving compression rate of $1/r$, while integrated generates exactly a sort of coupling inducing the logarithmic spiral pattern. 

The three equations governing mass and momentum conservation form a partial differential system with cubic nonlinearity. A linearization procedure demonstrates the evidence the formation od linearly stable spiral patterns including Archimedean and logarithmic ones. In order to find fully nonlinear solutions we write the density field as a real magnitude and a phase function, and we demonstrate that with a sufficiently good physical approximation the dynamical equation for the phase can be mapped into a sine-Gordon equation whose stable multi-front solutions generate the ice patterns. By truncating the nonlinear system to its  quadratic terms  we obtain spiral solutions with logarithmic geometry. In addition to the existence of sine-Gordon solitons, the nonlinear solutions are analyzed using two additional  mathematical approaches: one predicting the formation of patterns as Townes solitary modes, and another using a series expansion.  Pure radial, azimuthal and spiral modes are obtained from the fully nonlinear equations. Combinations of multiple-spiral solutions are also obtained, matching the experimental observations. The  nonlinear stability of the spiral patterns is analyzed by Arnold's convexity method, and the Hamiltonian of the solutions is plotted versus some order parameters showing the existence of geometric phase transitions.

\vskip0.5cm

\textbf{Acknowledgements}
One of the authors (AL) is grateful to Dalian University of Technology  for sharing the preliminary results on this project, and for funding and hospitality during the accomplishment of this research project in 2018-2019.

\end{document}